\documentclass[aps,prb,twocolumn,showpacs,amsmath]{revtex4}
\usepackage{latexsym}
\usepackage{amssymb}
\usepackage{bm}
\usepackage{graphicx}

\newcommand{\pdag}{{\phantom{\dagger}}}
\newcommand{\bq}{\begin{equation}}
\newcommand{\eq}{\end{equation}}
\newcommand{\bn}{\begin{eqnarray}}
\newcommand{\en}{\end{eqnarray}}

\begin{document}

\title{First-order coherent resonant tunneling through an interacting coupled-quantum-dot 
interferometer: generic quantum rate equations and current noise}

\author{Bing Dong} 
\affiliation{Department of Physics, Shanghai Jiaotong University,
1954 Huashan Road, Shanghai 200030, China}

\author{X. L. Lei} 
\affiliation{Department of Physics, Shanghai Jiaotong University,
1954 Huashan Road, Shanghai 200030, China}

\author{Norman J. M. Horing}
\affiliation{Department of Physics and Engineering Physics, Stevens
Institute of Technology, Hoboken, New Jersey 07030, USA}

\begin{abstract}

We carry out a detailed analysis of coherent resonant tunneling through two coupled quantum dots 
(CQD) in a parallel arrangement in the weak tunneling limit. We establish a set of quantum rate 
equations (QREs) in terms of the eigenstate-representation by means of a generic quantum Langevin 
equation approach, which is valid for arbitrary bias-voltage, temperature, and interdot hopping 
strength. Based on linear-response theory, we further derive the current and 
frequency-independent shot noise formulae. Our results reveal that a previously used formula for 
evaluating Schottky-type noise of a ``classical" single-electron transistor is a direct result of 
linear-response theory, and it remains applicable for small quantum devices with internal 
coupling. Our numerical calculations show some interesting transport features (i) for a 
series-CQD: the appearance of a NDC due to the bias-voltage-induced shifting of bare levels or a 
finite interdot Coulomb repulsion, and (ii) for a parallel CQD in strong interdot Coulomb 
repulsion regime: finite-bias-induced AB oscillations of current, and magnetic-flux-controllable 
negative differential conductance and a huge Fano factor.        

\end{abstract}


\maketitle

\section{Introduction}

The investigation of quantum oscillations between two levels in a coupled quantum dot (CQD) 
system under transport conditions has been the subject of enormous interest over the last 
decade.\cite{Haug,Wiel} 
Recently, a CQD arranged in parallel between source and drain has been experimentally reported to 
constitute a mesoscopic QD Aharonov-Bohm (AB) interferometer,\cite{Holleitner1,Holleitner2,ChenJ} 
and it is believed that manipulation of each of the QDs separately and the application of 
magnetic-flux piercing the device can provide controllable parameters for the design of transport 
properties by tuning quantum 
oscillations.\cite{Holleitner1,Holleitner2,ChenJ,Loss,Sukhorukov,Lev,Konig,Kub
ala,Shahbazyan,Ladron,Dong1}

To describe such quantum oscillations in quantum transport, master equations and a ``quantum" 
version of rate equations were first proposed by Nazarov and coworkers,\cite{Nazarov,Stoof} and 
later derived from the Schr\"odinger equation,\cite{Gurvitz,Gurvitz1} respectively. These 
original works are mainly for a CQD in a series arrangement between leads in the limit of zero 
temperature and large bias-voltage. The authors have generalized the quantum rate equations 
(QREs) for the case of a CQD interferometer in the same limit and employed them to study 
magnetic-flux-controlled photon-assisted tunneling.\cite{Dong1} 
Recently, a bias-voltage- and temperature-dependent generalization of the QREs was carried out by 
the authors employing the nonequilibrium Green's function in the limit of weak dot-dot 
coupling.\cite{Dong} However, the dot-dot coupling of the CQD device is usually tuned using gate 
voltage in experiments, and it may not be weaker than the tunnel-coupling to external electrodes. 
Therefore, it is desirable to develop generic QREs without assuming weakness of the dot-dot 
coupling for the purpose of systematically analyzing the transport properties of a CQD system. 

To accomplish this, we employ a generic quantum Langevin equation approach to derive the QREs in 
the eigenstate representation. The main result is the determination of a new coherent transfer 
term that emerges in all dynamic equations of the reduced density-matrix (RDM) elements stemming 
from the effective coupling between two eigenstates due to tunneling, which is absent from our 
previous results for QREs.\cite{Dong} In addition, other coherent terms occur in the ensuing QREs 
for a CQD interferometer due to interference between the two path branches enclosing 
magnetic-flux, which is responsible for the AB oscillation feature of transport, as expected.            

In another emerging aspect of CQD systems, studies of current fluctuations have become an 
important topic.\cite{Blanter,Beenakker} Several analyses of shot noise in a CQD system have been 
undertaken by means of the QRE 
approach.\cite{Sun,Elattari,Aguado,Djuric,Djuric2,Kieblich,Kieblich1,Aghassi,D
ong4} Most of these works have involved calculations in the large bias-voltage limit for CQDs in 
series\cite{Sun,Elattari,Aguado,Djuric2,Kieblich} or in parallel.\cite{Dong4} Our earlier work 
was carried out with analyses based on our previous QREs, and, correspondingly it is only valid 
for the case of weak dot-dot hopping.\cite{Djuric2} Moreover, some other recent studies of 
bias-voltage-dependent shot noise of a CQD have not treated the quantum coherence 
effect.\cite{Kieblich1,Aghassi} Therefore, in this paper, we will also analyze zero-frequency 
shot noise of an interacting CQD using the presently developed QREs, focusing our attention on 
the coherence and interference effects and its magnetic-flux dependence at both large and small 
bias-voltages.    

The outline of the paper is as follows. In Sec.~II, we present our physical model for a CQD 
interferometer connected to two electrodes and derive a set of QREs using a generic quantum 
Langevin equation approach with a Markovian approximation in describing the dynamic evolution of 
the RDM elements under transport conditions. Such microscopically derived QREs are valid at 
arbitrary bias-voltage and temperature, and dot-dot hopping strength as well. In this section, we 
also derive the current and frequency-independent shot noise formulae in terms of the RDM 
elements using linear-response theory. Employing the obtained formulae, we then investigate the 
transport properties of a CQD in series in Sec.~III, stressing the asymmetric transport property. 
Moreover, we study in detail the combined effect of the additional coherent transfer term and the 
two-pathway-interference term on current (Sec.~IV) and on shot noise (Sec.~V) of a CQD 
interferometer.   
A summary is given in Sec.~VI.

\section{Hamiltonian and Formulation}

The parallel-coupled interacting QD interferometer connected to two normal leads is illustrated 
in Fig.~1. 
The Hamiltonian is given by
\bq
H=H_{L}+H_{R}+H_{D}+H_{T}, \label{ham}
\eq    
where $H_{\eta}$ ($\eta=L,R$) describes noninteracting electron baths in the left, right leads, 
respectively:
\bq
H_{\eta} = \sum_{{\bf k}} (\varepsilon_{\eta{\bf k}} - \mu_{\eta}) a_{\eta {\bf k}}^\dagger 
a_{\eta{\bf k}}^\pdag,
\eq
with $a_{\eta{\bf k}}$ being the annihilation operator of an electron with momentum ${\bf k}$, 
and energy $\varepsilon_{\eta {\bf k}}$ in lead $\eta$, and $\mu_{\eta}$ being the effective 
Fermi energy of lead $\eta$. In our studies, a bias-voltage $V$ is taken to be applied 
symmetrically between the two electrodes, i.e., $\mu_{L}=-\mu_{R}=eV/2$, to sustain a persistent 
electron flow from one lead to the other. In equilibrium, $V=0$, we set $\mu_L=\mu_R=0$. 
Throughout, we will use units with $\hbar=k_B=e=1$.    

\begin{figure}[htb]
\includegraphics[height=4cm,width=8cm]{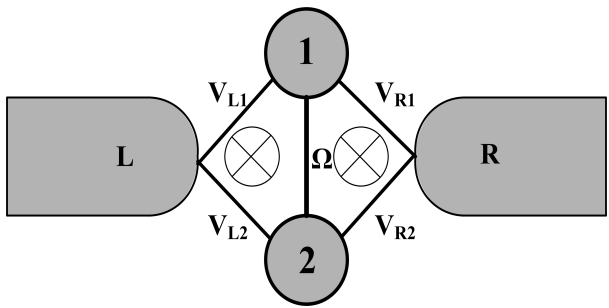}
\caption{Schematic diagram for coherent resonant tunneling through a parallel-coupled double 
quantum dot system in an Aharonov-Bohm interferometer.}
\label{fig1}
\end{figure}

The Hamiltonian of the isolated CQD system, $H_D$, is:
\bq
H_{D}=\sum_{j=1,2} \varepsilon_{j}c_{j}^\dagger c_{j}^\pdag + U c_{1}^\dagger c_{1}^\pdag 
c_{2}^\dagger c_{2}^\pdag + \Omega ( c_{1}^\dagger c_{2}^\pdag + c_{2}^\dagger c_{1}^\pdag), 
\label{Hqd}
\eq    
where $c_{j}$ is the annihilation operator for a spinless electron in the $j$th QD ($j=1,2$). 
$\varepsilon_j$ is the energy of the single level in the $j$th QD, measured from the Fermi energy 
of the two electrodes at equilibrium, $\varepsilon_{1(2)}=\varepsilon_d \pm \delta$, with 
$\delta$ being the bare mismatch between the two bare levels. Here, we assume that only one 
single electron level in each dot contributes to current. 
It should be noted that since we consider first-order resonant tunneling in this paper, and take 
no account of spin-flip scattering and spin-flip cotunneling processes (second-order tunneling 
processes), it is reasonable to assume spinless electrons in our model.  
The second term represents the interdot Coulomb interaction $U$. The last term of Eq.~(\ref{Hqd}) 
denotes hopping, $\Omega$, between the two QDs.

In this CQD system, there are a total of four possible states for the present system: (1) the 
whole system is empty, $|0\rangle\equiv|0\rangle_{1} |0\rangle_{2}$, and its energy is zero; (2) 
the first QD is singly occupied by an electron, $|1\rangle\equiv|1 \rangle_{1} |0\rangle_{2}$, 
and its energy is $\varepsilon_{1}$; (3) the second QD is singly occupied, 
$|2\rangle\equiv|0\rangle_{1} |1\rangle_{2}$, and its energy is $\varepsilon_2$; (4) both dots 
are occupied, $|d\rangle\equiv |1\rangle_{1} |1\rangle_2$, and its energy is 
$\varepsilon_1+\varepsilon_2+U$. Furthermore, with the four possible single electronic states 
considered as the basis, the density-matrix elements may be expressed as $\hat\rho_{00}=|0\rangle 
\langle 0|$, $\hat\rho_{11}=|1\rangle\langle 1|$, $\hat\rho_{22}=|2\rangle\langle 2|$, 
$\hat\rho_{dd}=|d\rangle\langle d|$, and $\hat\rho_{12}=|1\rangle \langle 2|$. The statistical 
expectation values of the diagonal elements of the density-matrix, $\rho_{00}=\langle 
\hat\rho_{00}\rangle$, $\rho_{jj}=\langle \hat\rho_{jj}\rangle$ ($j=1,2$), and $\rho_{dd}=\langle 
\hat \rho_{dd} \rangle$, give the occupation probabilities of the electronic levels in the system 
being empty, or singly occupied in the $j$th QD by an electron, or doubly-occupied by electrons, 
respectively. The off-diagonal term $\rho_{12}=\langle \hat\rho_{12}\rangle$ describes coherent 
superposition involving two electronic occupation states, $|1 \rangle_{1} |0\rangle_{2}$ and 
$|0\rangle_{1} |1\rangle_{2}$.      

To properly account for interference effects between the two pathways for tunneling through the 
system at hand, it is convenient to diagonalize the Hamiltonian of the isolated CQD by a unitary 
transformation
\begin{subequations}
\label{transform}
\bn
|\alpha\rangle &=& \cos \frac{\theta}{2} |1\rangle + \sin \frac{\theta}{2} |2\rangle, \\
|\beta\rangle &=& \sin \frac{\theta}{2} |1\rangle - \cos \frac{\theta}{2} |2\rangle, \quad 
\theta = \arctan \frac{\Omega}{\delta}. 
\en
\end{subequations}
The transformed Hamiltonian is
\begin{subequations}
\bq
H_{D}=\lambda_{\alpha} |\alpha\rangle \langle \alpha |+ \lambda_{\beta} |\beta \rangle \langle 
\beta| + (\varepsilon_1+\varepsilon_2+U) |d\rangle\langle d|, \label{Hqdeigen}
\eq
where $\lambda_{\alpha(\beta)}$ is the eigen-energy, 
\bq
\lambda_{\alpha(\beta)}=\varepsilon_d \pm \Delta, \label{eigenergy}
\eq
\end{subequations}
with $\Delta=\sqrt{\Omega^2+ \delta^2}$.
Correspondingly, the density-matrix elements in the new double-dot eigenstate basis, 
$\hat\rho_{\chi \chi}=|\chi \rangle \langle \chi|$ and $\hat\rho_{\alpha \beta}= |\alpha \rangle 
\langle \beta|$, have similar physical meanings to those in the site-representation (SR). The 
relations between these density-matrix elements of singly-occupied states in different bases can 
easily be deduced from the unitary transformation, Eq.~(\ref{transform}).  
(Note that the empty state, $\hat\rho_{00}$, and the double occupation state, $\hat\rho_{dd}$, 
have the same meaning in the both representations.) 

The tunnel-coupling between the interferometer and the electrodes, $H_T$, can be written in the 
eigenstate-representation (ER) as ($s_\theta=\sin \frac{\theta}{2}$ and $c_\theta=\cos 
\frac{\theta}{2}$)
\bn
H_T &=& \sum_{{\bf k}} [(V_{L1} e^{i\varphi /4} a_{L {\bf k} }^{\dagger } + V_{R1} e^{-i\varphi 
/4} a_{R {\bf k} }^{\dagger } ) \cr
&& \times (c_\theta |0 \rangle \langle \alpha | + s_\theta |0 \rangle \langle \beta| - s_\theta 
|\alpha\rangle \langle d| + c_\theta |\beta \rangle \langle d|) \cr
&& + (V_{L2} e^{-i\varphi /4} a_{L {\bf k} }^{\dagger } + V_{R2} e^{i\varphi /4} a_{R {\bf k} 
}^{\dagger } ) \cr
&& \hspace{-1cm} \times ( s_\theta |0 \rangle \langle \alpha | - c_\theta |0 \rangle \langle 
\beta| + c_\theta |\alpha\rangle \langle d| + s_\theta |\beta\rangle \langle d|) +{\rm {H.c.}}]. 
\label{hamiltonianT}
\en
For simplicity, the lead-dot tunneling matrix elements, $V_{\eta j}$, are assumed to be real and 
independent of energy. The factor $e^{\pm i\varphi /4}$ is the accumulated Peierls phase due to 
the magnetic-flux $\Phi$ ($\varphi\equiv 2\pi \Phi/\Phi_0$, $\Phi_0\equiv hc/e$ is the 
magnetic-flux quantum) which penetrates the area enclosed by two tunneling pathways of the 
interferometer.

In the following, we apply a generic quantum Langevin equation 
approach\cite{Schwinger,Ackerhalt,Cohen,Milonni,Gardiner,Smirnov,Dong2} to derive a set of 
quantum rate equations (QREs) to describe the dynamics of the system variables of the CQD due to 
coherent resonant tunneling between external reservoirs, as modeled by Eq.~(\ref{ham}).
In the derivation, three steps are involved: First, we start from the Heisenberg equations of 
motion (EOMs) for the density-matrix operators $\hat \rho_{00}$, $\hat \rho_{\chi\chi'}$, and 
$\rho_{dd}$ in the ER and related reservoir operators $c_{\eta {\bf k}}$, and then formally solve 
them by integration of these EOMs exactly. Next, under the assumption that the time scale of 
decay processes is much slower than that of free evolution, which is reasonable in the 
weak-tunneling approximation, we replace the time-dependent operators involved in the integrals 
of these EOMs approximately in terms of their free evolutions. Thirdly, these EOMs are expanded 
in powers of the tunnel-coupling matrix element $V_{\eta j}$ up to second order. By adopting a 
Markovian approximation, we finally derive the generic QREs with arbitrary bias-voltage and 
temperature, as well as arbitrary dot-dot hopping, as
\begin{subequations}
\bn
\dot \rho_{00} &=& - [ c_{\theta}^2 (\Gamma_{11\alpha}^+ + \Gamma_{22\beta}^+) + s_{\theta}^2 ( 
\Gamma_{11\beta}^+ + \Gamma_{22\alpha}^+) \cr
&& +\frac{1}{2} \sin \theta (\Gamma_{12\alpha}^+ + \Gamma_{21\alpha}^+ - \Gamma_{12\beta}^+ 
-\Gamma_{21\beta}^+) ] \rho_{00} \cr
&& + [ c_{\theta}^2 \Gamma_{11\alpha}^- + s_{\theta}^2 \Gamma_{22\alpha}^- + \frac{1}{2} \sin 
\theta ( \Gamma_{12\alpha}^- + \Gamma_{21\alpha}^-)] \rho_{\alpha\alpha} \cr
&& + [ s_{\theta}^2 \Gamma_{11\beta}^- + c_{\theta}^2 \Gamma_{22\beta}^- - \frac{1}{2} \sin 
\theta (\Gamma_{12\beta}^- + \Gamma_{21\beta}^-)] \rho_{\beta\beta} \cr
&& + \frac{1}{4} \sin \theta (\Gamma_{11\alpha}^- + \Gamma_{11\beta}^- - \Gamma_{22\alpha}^- - 
\Gamma_{22\beta}^-) (\rho_{\alpha\beta} + \rho_{\beta\alpha}) \cr
&& + \frac{1}{2}[s_{\theta}^2 (\Gamma_{12\alpha}^- + \Gamma_{12\beta}^-) - c_{\theta}^2 
(\Gamma_{21\alpha}^- + \Gamma_{21\beta}^-)] \rho_{\alpha\beta} \cr  
&& - \frac{1}{2}[c_{\theta}^2 (\Gamma_{12\alpha}^- + \Gamma_{12\beta}^-) - s_{\theta}^2 
(\Gamma_{21\alpha}^- + \Gamma_{21\beta}^-)] \rho_{\beta\alpha}, \label{qrer00} 
\en
\bn
\dot \rho_{\alpha\alpha} &=& [c_{\theta}^2 \Gamma_{11\alpha}^+ + s_{\theta}^2 \Gamma_{22\alpha}^+ 
+ \frac{1}{2} \sin \theta ( \Gamma_{12\alpha}^+ + \Gamma_{21\alpha}^+)] \rho_{00} \cr
&& - [c_{\theta}^2 \Gamma_{11\alpha}^- + s_{\theta}^2 \Gamma_{22\alpha}^- + \frac{1}{2} \sin 
\theta (\Gamma_{12\alpha}^- + \Gamma_{21\alpha}^-)] \rho_{\alpha\alpha} \cr
&& -[s_{\theta}^2 \widetilde{\Gamma}_{11\beta}^+ + c_{\theta}^2 \widetilde{\Gamma}_{22\beta}^+ - 
\frac{1}{2}\sin \theta (\widetilde{\Gamma}_{12\beta}^+ + \widetilde{\Gamma}_{21\beta}^+)] 
\rho_{\alpha\alpha} \cr
&& + [c_{\theta}^2 \widetilde{\Gamma}_{22\beta}^- + s_{\theta}^2 \widetilde{\Gamma}_{11\beta}^- - 
\frac{1}{2} \sin \theta (\widetilde{\Gamma}_{12\beta}^- + \widetilde{\Gamma}_{21\beta}^-)] 
\rho_{dd} \cr
&& - \frac{1}{4} \sin \theta (\Gamma_{11\beta}^- - \Gamma_{22\beta}^- - 
\widetilde{\Gamma}_{11\alpha}^+ + \widetilde{\Gamma}_{22\alpha}^+) (\rho_{\alpha\beta} + 
\rho_{\beta\alpha}) \cr
&& - \frac{1}{2} [s_{\theta}^2 (\Gamma_{12\beta}^- - \widetilde{\Gamma}_{12\alpha}^+) - 
c_{\theta}^2 (\Gamma_{21\beta}^- -\widetilde{\Gamma}_{21\alpha}^+)] \rho_{\alpha\beta} \cr
&& - \frac{1}{2} [s_{\theta}^2 (\Gamma_{21\beta}^- - \widetilde{\Gamma}_{21\alpha}^+) - 
c_{\theta}^2 (\Gamma_{12\beta}^- - \widetilde{\Gamma}_{12\alpha}^+)] \rho_{\beta\alpha} , 
\label{qreraa}
\en
\bn
\dot \rho_{\beta\beta} &=& [s_{\theta}^2 \Gamma_{11\beta}^+ + c_{\theta}^2 \Gamma_{22\beta}^+ - 
\frac{1}{2} \sin \theta ( \Gamma_{12\beta}^+ + \Gamma_{21\beta}^+)] \rho_{00} \cr
&& - [s_{\theta}^2 \Gamma_{11\beta}^- + c_{\theta}^2 \Gamma_{22\beta}^- - \frac{1}{2} \sin \theta 
(\Gamma_{12\beta}^- + \Gamma_{21\beta}^-)] \rho_{\beta\beta} \cr
&& -[c_{\theta}^2 \widetilde{\Gamma}_{11\alpha}^+ + s_{\theta}^2 \widetilde{\Gamma}_{22\alpha}^+ 
+ \frac{1}{2}\sin \theta (\widetilde{\Gamma}_{12\alpha}^+ + \widetilde{\Gamma}_{21\alpha}^+)] 
\rho_{\beta\beta} \cr
&& + [s_{\theta}^2 \widetilde{\Gamma}_{22\alpha}^- + c_{\theta}^2 \widetilde{\Gamma}_{11\alpha}^- 
+ \frac{1}{2} \sin \theta (\widetilde{\Gamma}_{12\alpha}^- + \widetilde{\Gamma}_{21\alpha}^-)] 
\rho_{dd} \cr
&& - \frac{1}{4} \sin \theta (\Gamma_{11\alpha}^- - \Gamma_{22\alpha}^- - 
\widetilde{\Gamma}_{11\beta}^+ + \widetilde{\Gamma}_{22\beta}^+) (\rho_{\alpha\beta} + 
\rho_{\beta\alpha}) \cr
&& - \frac{1}{2} [s_{\theta}^2 (\Gamma_{12\alpha}^- -\widetilde{\Gamma}_{12\beta}^+) - 
c_{\theta}^2 (\Gamma_{21\alpha}^- - \widetilde{\Gamma}_{21\beta}^+)] \rho_{\alpha\beta} \cr
&& - \frac{1}{2} [s_{\theta}^2 (\Gamma_{21\alpha}^- - \widetilde{\Gamma}_{21\beta}^+) - 
c_{\theta}^2 (\Gamma_{12\alpha}^- -\widetilde{\Gamma}_{12\beta}^+)] \rho_{\beta\alpha} ,
\en
\bn
\dot \rho_{dd} &=& [s_{\theta}^2 \widetilde{\Gamma}_{11\beta}^+ + c_{\theta}^2 
\widetilde{\Gamma}_{22\beta}^+ - \frac{1}{2} \sin (\widetilde{\Gamma}_{12\beta}^+ + 
\widetilde{\Gamma}_{21\beta}^+)] \rho_{\alpha\alpha} \cr
&& + [c_{\theta}^2 \widetilde{\Gamma}_{11\alpha}^+ + s_{\theta}^2 \widetilde{\Gamma}_{22\alpha}^+ 
+ \frac{1}{2} \sin (\widetilde{\Gamma}_{12\alpha}^+ + \widetilde{\Gamma}_{21\alpha}^+)] 
\rho_{\beta\beta} \cr
&& - [s_{\theta}^2 (\widetilde{\Gamma}_{11\beta}^- + \widetilde{\Gamma}_{22\alpha}^-) + 
c_{\theta}^2 (\widetilde{\Gamma}_{11\alpha}^- + \widetilde{\Gamma}_{22\beta}^-) \cr
&& + \frac{1}{2} \sin \theta (\widetilde{\Gamma}_{12\alpha}^- - \widetilde{\Gamma}_{12\beta}^- + 
\widetilde{\Gamma}_{21\alpha}^- - \widetilde{\Gamma}_{21\beta}^-)] \rho_{dd} \cr
&& -\frac{1}{4} \sin \theta (\widetilde{\Gamma}_{11\alpha}^+ + \widetilde{\Gamma}_{11\beta}^+ - 
\widetilde{\Gamma}_{22\alpha}^+ - \widetilde{\Gamma}_{22\beta}^+) (\rho_{\alpha\beta}+ 
\rho_{\beta\alpha}) \cr
&& - \frac{1}{2} [s_{\theta}^2 (\widetilde{\Gamma}_{12\alpha}^+ + \widetilde{\Gamma}_{12\beta}^+) 
- c_{\theta}^2 (\widetilde{\Gamma}_{21\alpha}^+ + \widetilde{\Gamma}_{21\beta}^+)] 
\rho_{\alpha\beta} \cr
&& + \frac{1}{2} [c_{\theta}^2 (\widetilde{\Gamma}_{12\alpha}^+ + \widetilde{\Gamma}_{12\beta}^+) 
- s_{\theta}^2 (\widetilde{\Gamma}_{21\alpha}^+ + \widetilde{\Gamma}_{21\beta}^+)] 
\rho_{\beta\alpha},
\en 
\bn
\dot \rho_{\alpha\beta} &=& i2\Delta \rho_{\alpha\beta} + [ \frac{1}{4} \sin \theta 
(\Gamma_{11\alpha}^+ + \Gamma_{11\beta}^+ - \Gamma_{22\alpha}^+ - \Gamma_{22\beta}^+) \cr
&& - \frac{1}{2} c_{\theta}^2 (\Gamma_{12\alpha}^+ + \Gamma_{12\beta}^+) + \frac{1}{2} 
s_{\theta}^2 (\Gamma_{21\alpha}^+ + \Gamma_{21\beta}^+)] \rho_{00} \cr
&& -\frac{1}{2} [ c_{\theta}^2 (\Gamma_{11\alpha}^- + \Gamma_{22\beta}^- + 
\widetilde{\Gamma}_{11\alpha}^+ + \widetilde{\Gamma}_{22\beta}^+) \cr
&& + s_{\theta}^2 (\Gamma_{11\beta}^- + \Gamma_{22\alpha}^- + \widetilde{\Gamma}_{11\beta}^+ + 
\widetilde{\Gamma}_{22\alpha}^+) \cr
&& + \frac{1}{2} \sin \theta (\Gamma_{12\alpha}^- + \Gamma_{21\alpha}^- - \Gamma_{12\beta}^- - 
\Gamma_{21\beta}^- \cr
&& + \widetilde{\Gamma}_{12\alpha}^+ + \widetilde{\Gamma}_{21\alpha}^+ - 
\widetilde{\Gamma}_{12\beta}^+ - \widetilde{\Gamma}_{21\beta}^+) ] \rho_{\alpha\beta} \cr
&& + \frac{1}{2} [ c_{\theta}^2 (\Gamma_{12\alpha}^- -\widetilde{\Gamma}_{12\beta}^+) - 
s_{\theta}^2 (\Gamma_{21\alpha}^- - \widetilde{\Gamma}_{21\beta}^+) \cr
&& - \frac{1}{2} \sin \theta (\Gamma_{11\alpha}^- - \Gamma_{22\alpha}^- 
-\widetilde{\Gamma}_{11\beta}^+ + \widetilde{\Gamma}_{22\beta}^+)] \rho_{\alpha\alpha} \cr
&& + \frac{1}{2} [ c_{\theta}^2 (\Gamma_{12\beta}^- -\widetilde{\Gamma}_{12\alpha}^+) - 
s_{\theta}^2 (\Gamma_{21\beta}^- -\widetilde{\Gamma}_{21\alpha}^+) \cr
&& - \frac{1}{2} \sin \theta (\Gamma_{11\beta}^- - \Gamma_{22\beta}^- - 
\widetilde{\Gamma}_{11\alpha}^+ + \widetilde{\Gamma}_{22\alpha}^+)] \rho_{\beta\beta} \cr
&& + \frac{1}{2} [\frac{1}{2} \sin \theta (\widetilde{\Gamma}_{22\alpha}^- + 
\widetilde{\Gamma}_{22\beta}^- - \widetilde{\Gamma}_{11\alpha}^- - 
\widetilde{\Gamma}_{11\beta}^-) \cr
&& + c_{\theta}^2 (\widetilde{\Gamma}_{12\alpha}^- + \widetilde{\Gamma}_{12\beta}^-) - 
s_{\theta}^2 (\widetilde{\Gamma}_{21\alpha}^- + \widetilde{\Gamma}_{21\beta}^-)] \rho_{dd},\cr
&& \label{qrerab}
\en
\end{subequations}
where the electron tunneling-in(out) rates are defined as
\begin{subequations}
\bn
\Gamma_{jj\chi}^\pm &=& \sum_{\eta}\Gamma_{\eta jj \chi}^\pm = \sum_{\eta} \Gamma_{\eta jj} 
f_{\eta}^\pm(\lambda_{\chi}),\label{tr1} \\
\Gamma_{12\chi}^\pm &=& \sum_{\eta}\Gamma_{\eta 12 \chi}^\pm = \sum_{\eta} \Gamma_{\eta 12} 
e^{s_{\eta} i \varphi/2} f_{\eta}^\pm(\lambda_{\chi}),\\
\Gamma_{21\chi}^\pm &=& \sum_{\eta}\Gamma_{\eta 21 \chi}^\pm = \sum_{\eta} \Gamma_{\eta 12} 
e^{-s_{\eta} i \varphi/2} f_{\eta}^\pm(\lambda_{\chi}),\cr
&& \label{tr3} \\
\widetilde{\Gamma}_{jj\chi}^\pm &=& \sum_{\eta}\widetilde{\Gamma}_{\eta jj \chi}^\pm = 
\sum_{\eta} \Gamma_{\eta jj} f_{\eta}^\pm(\lambda_{\chi}+U),\label{tr1u} \\
\widetilde{\Gamma}_{12\chi}^\pm &=& \sum_{\eta}\widetilde{\Gamma}_{\eta 12 \chi}^\pm = 
\sum_{\eta} \Gamma_{\eta 12} e^{s_{\eta} i \varphi/2} f_{\eta}^\pm(\lambda_{\chi}+U),\\
\widetilde{\Gamma}_{21\chi}^\pm &=& \sum_{\eta}\widetilde{\Gamma}_{\eta 21 \chi}^\pm = 
\sum_{\eta} \Gamma_{\eta 12} e^{-s_{\eta} i \varphi/2} f_{\eta}^\pm(\lambda_{\chi}+U),\cr
&& \label{tr3u}
\en
\end{subequations}
[$f_\eta^+(\epsilon)$ is the Fermi distribution function of lead $\eta$,  
$f_{\eta}^-(\epsilon)=1-f_{\eta}^+(\epsilon)$, and $s_{L/R}=\pm 1$] with
\bq
\Gamma_{\eta jj'}= 2\pi \varrho_{\eta} V_{\eta j}V_{\eta j'},
\eq
being constant in the wide band limit ($\varrho_{\eta}$ is the density of states of lead $\eta$). 
$\Gamma_{\eta jj'}$ represents the strength of tunnel-coupling between the $j$th QD level and 
lead $\eta$ if $j=j'$, and otherwise ($j\neq j'$) it measures the interference in tunneling 
through the different pathways. 
The adjoint equation of Eq.~(\ref{qrerab}) gives the equation of motion for the RDM off-diagonal 
element $\rho_{\beta\alpha}$. Also, we must note the normalization relation $\rho_{00}+ 
\sum_{\chi} \rho_{\chi\chi}+ \rho_{dd}=1$. Simple rate equations written in the ER were obtained 
earlier in Ref.~\onlinecite{Korotkov} for double-well semiconductor heterostructures.

In the ER, the CQD interferometer is equivalent to a single QD with two energy-levels, both of 
which couple to the two electrodes such that either level is tunnel-coupled to either electrode 
via two different pathways with different tunneling matrix elements depending on differing 
accumulated phases due to the magnetic-flux [see Eq.~(\ref{hamiltonianT})]. Figure 
\ref{fig:energycon} below exhibits schematic diagrams of energy configurations for coherent 
tunneling through a CQD in the ER. In this situation, all tunneling events can be classified in 
three different categories. For example, $\Gamma_{11\alpha}^{+}$ ($\Gamma_{11\alpha}^{-}$) 
describes the tunneling rate of an electron entering (leaving) level $\alpha$ in the CQD without 
the occupancy of level $\beta$ via pathway $1$ (indicated by the upper line in Fig.~1); 
Similarly, $\Gamma_{22\alpha}^{+}$ ($\Gamma_{22\alpha}^{-}$) denotes the same tunneling process 
via pathway $2$ (indicated by the lower line in Fig.~1); $\Gamma_{12\alpha}^{+}$ 
($\Gamma_{12\alpha}^{-}$) describes the interferential tunneling process of an electron entering 
(leaving) level $\alpha$ without the occupancy of level $\beta$ due to the interference between 
the two pathways $1$ and $2$. All $\widetilde{\Gamma}_{jj'\chi}^{\pm}$ are the corresponding 
rates in the case of double occupation. Naturally, only the interferential tunneling term suffers 
an accumulated phase factor as shown in Eqs.~(\ref{tr3}) and (\ref{tr3u}).

Therefore, the classical parts of the dynamical equations of the RDM diagonal elements have clear 
classical interpretations. For example, the rate of change of electron number in level $\alpha$, 
$\rho_{\alpha\alpha}$, in the CQD, governed by Eq.~(\ref{qreraa}), is contributed to by the 
following four single-particle tunneling processes: (1) tunneling into level $\alpha$ of the CQD 
from both left and right leads, if the CQD is initially in the empty state $\rho_{00}$; (2) 
tunneling out from(into) level $\alpha$($\beta$) of the CQD into(from) both leads, when the CQD 
is initially just in the state $\rho_{\alpha\alpha}$($\rho_{\beta\beta}$); (3) tunneling out from 
level $\beta$ of the CQD into both leads, when the CQD is initially in the state $\rho_{dd}$, via 
path $1$, path $2$ and interference contributions. The last three terms in Eq.~(\ref{qreraa}) 
describe transitions between two eigenstates through the effective coupling with off-diagonal 
elements via both paths and interference, which have no classical counterpart. They are 
responsible for coherent (quantum) effects in the transport.     
We note that even though there is no direct coupling between the two eigenstates of the isolated 
Hamiltonian, Eq.~(\ref{Hqdeigen}), in the ER, the tunnel-coupling described by 
Eq.~(\ref{hamiltonianT}) results in an effective transition between them and thus leads to a 
quantum superposition state between the two eigenstates, whose dynamics is ruled by the equation 
of motion of the off-diagonal element $\rho_{\alpha\beta}$, Eq.~(\ref{qrerab}). 

The tunneling current operator through the interferometer is defined as the time rate of change 
of the charge density, $N_{\eta}=\sum_{{\bf k}} a_{\eta {\bf k} }^{\dagger} a_{\eta {\bf k} 
}^{\pdag}$, in lead $\eta$:
\bq
J_{\eta}= - \dot N_{\eta} = i [N_{\eta}, H]. \label{i}
\eq
According to linear-response theory in the interaction picture, we have:\cite{Mahan}
\bn
I_L &=& \langle J_{L} \rangle =-i \int_{-\infty}^t dt' \langle [ J_{L}(t), H_{I}(t')]_- \rangle .
\en  
Along the same procedure as above, we arrive at the following expression for the current within 
the weak tunnel-coupling approximation,
\bn
I_L &=& [c_{\theta}^2 (\Gamma_{L11\alpha}^+ + \Gamma_{L22\beta}^+) + s_{\theta}^2 
(\Gamma_{L11\beta}^+ + \Gamma_{L22\alpha}^+) \cr
&& +\frac{1}{2} \sin \theta ( \Gamma_{L12\alpha}^+ + \Gamma_{L21\alpha}^+ - \Gamma_{L12\beta}^+ - 
\Gamma_{L21\beta}^+)] \rho_{00} \cr
&& \hspace{-1cm} - [c_{\theta}^2 \Gamma_{L11\alpha}^- + s_{\theta}^2 \Gamma_{L22\alpha}^- + 
\frac{1}{2} \sin \theta (\Gamma_{L12\alpha}^- + \Gamma_{L21\alpha}^-)] \rho_{\alpha\alpha} \cr
&& \hspace{-1cm} + [s_{\theta}^2 \widetilde{\Gamma}_{L11\beta}^+ + c_{\theta}^2 
\widetilde{\Gamma}_{L22\beta}^- - \frac{1}{2} \sin \theta (\widetilde{\Gamma}_{L12\beta}^+ + 
\widetilde{\Gamma}_{L21\beta}^-)] \rho_{\alpha\alpha} \cr
&& \hspace{-1cm} - [s_{\theta}^2 \Gamma_{L11\beta}^- + c_{\theta}^2 \Gamma_{L22\beta}^- - 
\frac{1}{2} \sin \theta (\Gamma_{L12\beta}^- + \Gamma_{L21\beta}^-)] \rho_{\beta\beta} \cr
&& \hspace{-1cm} + [c_{\theta}^2 \widetilde{\Gamma}_{L11\alpha}^- + s_{\theta}^2 
\widetilde{\Gamma}_{L22\alpha}^- + \frac{1}{2} \sin \theta (\widetilde{\Gamma}_{L12\alpha}^- + 
\widetilde{\Gamma}_{L21\alpha}^-)] \rho_{\beta\beta} \cr
&& - \frac{1}{2} [\frac{1}{2} \sin \theta (\Gamma_{L11\alpha}^- + \Gamma_{L11\beta}^- - 
\Gamma_{L22\alpha}^- - \Gamma_{L22\beta}^-) \cr
&& + s_{\theta}^2 (\Gamma_{L12\alpha}^- + \Gamma_{L12\beta}^-) - c_{\theta}^2 
(\Gamma_{L21\alpha}^- + \Gamma_{L21\beta}^-)] \rho_{\alpha\beta} \cr
&& + \frac{1}{2} [\frac{1}{2} \sin \theta (\widetilde{\Gamma}_{L22\alpha}^+ + \Gamma_{L22\beta}^+ 
- \widetilde{\Gamma}_{L11\alpha}^+ - \widetilde{\Gamma}_{L11\beta}^+) \cr
&& - s_{\theta}^2 (\widetilde{\Gamma}_{L12\alpha}^+ + \widetilde{\Gamma}_{L12\beta}^+) + 
c_{\theta}^2 (\widetilde{\Gamma}_{L21\alpha}^+ + \widetilde{\Gamma}_{L21\beta}^+)] 
\rho_{\alpha\beta} \cr
&& - \frac{1}{2} [\frac{1}{2} \sin \theta (\Gamma_{L11\alpha}^- + \Gamma_{L11\beta}^- - 
\Gamma_{L22\alpha}^- - \Gamma_{L22\beta}^-) \cr
&& - c_{\theta}^2 (\Gamma_{L12\alpha}^- + \Gamma_{L12\beta}^-) + s_{\theta}^2 
(\Gamma_{L21\alpha}^- + \Gamma_{L21\beta}^-)] \rho_{\beta\alpha} \cr
&& + \frac{1}{2} [\frac{1}{2} \sin \theta (\widetilde{\Gamma}_{L22\alpha}^+ + 
\widetilde{\Gamma}_{L22\beta}^+ - \widetilde{\Gamma}_{L11\alpha}^+ - 
\widetilde{\Gamma}_{L11\beta}^+) \cr
&& + c_{\theta}^2 (\widetilde{\Gamma}_{L12\alpha}^+ + \widetilde{\Gamma}_{L12\beta}^+) - 
s_{\theta}^2 (\widetilde{\Gamma}_{L21\alpha}^+ + \widetilde{\Gamma}_{L21\beta}^+)] 
\rho_{\beta\alpha} \cr
&& - [s_{\theta}^2 \widetilde{\Gamma}_{L11\beta}^- + c_{\theta}^2 
\widetilde{\Gamma}_{L11\alpha}^- + s_{\theta}^2 \widetilde{\Gamma}_{L22\alpha}^- + c_{\theta}^2 
\widetilde{\Gamma}_{L22\beta}^- \cr
&& - \frac{1}{2} \sin \theta (\widetilde{\Gamma}_{L12\beta}^- - \widetilde{\Gamma}_{L12\alpha}^- 
+ \widetilde{\Gamma}_{L21\beta}^- - \widetilde{\Gamma}_{L21\alpha}^-)] \rho_{dd}. \cr
&& \label{currentL}
\en
Interchanging the subscripts ``L" and ``R", we obtain the tunneling current $I_R$ relevant to the 
right lead. It is easy to verify that $I_L=-I_R$.
This current formula demonstrates that all possible tunneling events relevant to lead $\eta$ 
(tunneling through both paths and with the interferential term) provide corresponding 
contributions to the current of lead $\eta$; the current is determined not only by the RDM 
diagonal elements, but it also involves the off-diagonal elements explicitly.

Another observable of interest is the current noise, whose spectrum is defined as the Fourier 
transform of the symmetric current-current correlation function
\bq
S_{\eta\eta'}(\omega)= \int_{-\infty}^\infty d\tau e^{i \omega \tau} \frac{1}{2} \langle [\delta 
J_{\eta}(t), \delta J_{\eta'}(t')]_+ \rangle,
\eq
with $\delta J_{\eta}(t)=J_{\eta}(t)- I_{\eta}$. Here, we only consider the frequency-independent 
(Schottky-type) shot noise. 
Employing linear-response theory we can also evaluate its statistical expectation value to the 
first nonvanishing term in $H_{I}$, obtaining   
\bn
S_{LL} &=& \frac{1}{2}\int_{-\infty}^{\infty} d\tau \langle [ J_{L}^o(t), J_{L}^o(t')]_+ \rangle 
\cr
&=& [c_{\theta}^2 (\Gamma_{L11\alpha}^+ + \Gamma_{L22\beta}^+) + s_{\theta}^2 
(\Gamma_{L11\beta}^+ + \Gamma_{L22\alpha}^+) \cr
&& +\frac{1}{2} \sin \theta ( \Gamma_{L12\alpha}^+ + \Gamma_{L21\alpha}^+ - \Gamma_{L12\beta}^+ - 
\Gamma_{L21\beta}^+)] \rho_{00} \cr
&& \hspace{-1cm} + [c_{\theta}^2 \Gamma_{L11\alpha}^- + s_{\theta}^2 \Gamma_{L22\alpha}^- + 
\frac{1}{2} \sin \theta (\Gamma_{L12\alpha}^- + \Gamma_{L21\alpha}^-)] \rho_{\alpha\alpha} \cr
&& \hspace{-1cm} + [s_{\theta}^2 \widetilde{\Gamma}_{L11\beta}^+ + c_{\theta}^2 
\widetilde{\Gamma}_{L22\beta}^- - \frac{1}{2} \sin \theta (\widetilde{\Gamma}_{L12\beta}^+ + 
\widetilde{\Gamma}_{L21\beta}^-)] \rho_{\alpha\alpha} \cr
&& \hspace{-1cm} + [s_{\theta}^2 \Gamma_{L11\beta}^- + c_{\theta}^2 \Gamma_{L22\beta}^- - 
\frac{1}{2} \sin \theta (\Gamma_{L12\beta}^- + \Gamma_{L21\beta}^-)] \rho_{\beta\beta} \cr
&& \hspace{-1cm} + [c_{\theta}^2 \widetilde{\Gamma}_{L11\alpha}^- + s_{\theta}^2 
\widetilde{\Gamma}_{L22\alpha}^- + \frac{1}{2} \sin \theta (\widetilde{\Gamma}_{L12\alpha}^- + 
\widetilde{\Gamma}_{L21\alpha}^-)] \rho_{\beta\beta} \cr
&& \hspace{-1cm} + \frac{1}{2} [\frac{1}{2} \sin \theta (\Gamma_{L11\alpha}^- + 
\Gamma_{L11\beta}^- - \Gamma_{L22\alpha}^- - \Gamma_{L22\beta}^-) \cr
&& \hspace{-1cm} + s_{\theta}^2 (\Gamma_{L12\alpha}^- + \Gamma_{L12\beta}^-) - c_{\theta}^2 
(\Gamma_{L21\alpha}^- + \Gamma_{L21\beta}^-)] \rho_{\alpha\beta} \cr
&& \hspace{-1cm} + \frac{1}{2} [\frac{1}{2} \sin \theta (\widetilde{\Gamma}_{L22\alpha}^+ + 
\Gamma_{L22\beta}^+ - \widetilde{\Gamma}_{L11\alpha}^+ - \widetilde{\Gamma}_{L11\beta}^+) \cr
&& \hspace{-1cm} - s_{\theta}^2 (\widetilde{\Gamma}_{L12\alpha}^+ + 
\widetilde{\Gamma}_{L12\beta}^+) + c_{\theta}^2 (\widetilde{\Gamma}_{L21\alpha}^+ + 
\widetilde{\Gamma}_{L21\beta}^+)] \rho_{\alpha\beta} \cr
&& \hspace{-1cm} + \frac{1}{2} [\frac{1}{2} \sin \theta (\Gamma_{L11\alpha}^- + 
\Gamma_{L11\beta}^- - \Gamma_{L22\alpha}^- - \Gamma_{L22\beta}^-) \cr
&& \hspace{-1cm} - c_{\theta}^2 (\Gamma_{L12\alpha}^- + \Gamma_{L12\beta}^-) + s_{\theta}^2 
(\Gamma_{L21\alpha}^- + \Gamma_{L21\beta}^-)] \rho_{\beta\alpha} \cr
&& \hspace{-1cm} + \frac{1}{2} [\frac{1}{2} \sin \theta (\widetilde{\Gamma}_{L22\alpha}^+ + 
\widetilde{\Gamma}_{L22\beta}^+ - \widetilde{\Gamma}_{L11\alpha}^+ - 
\widetilde{\Gamma}_{L11\beta}^+) \cr
&& \hspace{-1cm} + c_{\theta}^2 (\widetilde{\Gamma}_{L12\alpha}^+ + 
\widetilde{\Gamma}_{L12\beta}^+) - s_{\theta}^2 (\widetilde{\Gamma}_{L21\alpha}^+ + 
\widetilde{\Gamma}_{L21\beta}^+)] \rho_{\beta\alpha} \cr
&& \hspace{-1cm} + [s_{\theta}^2 \widetilde{\Gamma}_{L11\beta}^- + c_{\theta}^2 
\widetilde{\Gamma}_{L11\alpha}^- + s_{\theta}^2 \widetilde{\Gamma}_{L22\alpha}^- + c_{\theta}^2 
\widetilde{\Gamma}_{L22\beta}^- \cr
&& \hspace{-1cm} - \frac{1}{2} \sin \theta (\widetilde{\Gamma}_{L12\beta}^- - 
\widetilde{\Gamma}_{L12\alpha}^- + \widetilde{\Gamma}_{L21\beta}^- - 
\widetilde{\Gamma}_{L21\alpha}^-)] \rho_{dd}.
\label{SLL}
\en
The corresponding result for $S_{RR}$ is obtained by interchanging the subscripts ``L" and ``R" 
in Eq.~(\ref{SLL}). In leading order, $S_{LR}=S_{RL}=0$. 

This shot noise formula merits further discussion.
It is well-known that the Schottky noise originates from the self-correlation of a given 
tunneling event with itself. In literature, this term has been elaborately studied for a 
sequential picture of resonant tunneling through a double-barrier transistor using a master 
equation,\cite{Davies,Korotkov1} and also using a stochastic wave function 
approach.\cite{Wiseman} These previous studies show that the ``classical" intrinsic shot noise 
can be simply expressed by the formula\cite{Davies,Korotkov1}
\begin{subequations}
\bq
S_{\eta\eta}= \sum_{n} (+)\gamma_{\eta n} \bm \rho_n^0, \label{s-for}
\eq
provided the current can be written in the following form  
\bq
I_{\eta}=\sum_{n} (+/-)\gamma_{\eta n} \bm \rho_n^0, \label{i-for}
\eq
\end{subequations}
where $\bm \rho$ is a vector whose components are the RDM elements of the single-electron 
transistor, the superscript ``$0$" indicates its stationary solution based on the rate equation, 
and $\gamma_{\eta n}$ is the corresponding rate of a possible tunneling process that can change 
the state $n$ of the single-electron transistor and contribute to the current of lead $\eta$. 
This is to say that all single-particle tunneling processes involving lead $\eta$ will give a 
positive ($+$) or a negative ($-$) contribution to the current of lead $\eta$, depending on 
whether an electron is to enter or to leave this lead in a particular tunneling process; 
nevertheless self-correlation terms in all these same processes will always result in a positive 
($+$) contribution to the Schottky-type shot noise, and the only difference between the current 
and the self-correlation noise formulae, Eqs.~(\ref{i-for}) and (\ref{s-for}), is a sign change.  
We note that the formula Eq.~(\ref{s-for}) has been only proved for a ``classical" double-barrier 
tunneling device, in which the RDM of the device has only diagonal elements (thus $\bm 
\rho_n^0\geq 0$) and the tunneling rate $\gamma_{\eta n}$ is always non-negative ($\gamma_{\eta 
n}\geq 0$).   
Without attempting to rigorously verify its validity, this scheme has been recently applied 
directly to calculate the Schottky-type shot noise of a QD system having coupled internal degrees 
of freedom, but with its current being explicitly dependent only on RDM diagonal elements which 
always have positive values.\cite{Djuric,Djuric2} The present theoretical formulation circumvents 
this problem. Furthermore, we prove here that (1) this scheme is just a result of linear-response 
theory in the weak-tunneling limit; and (2) this scheme remains valid for the calculation of 
Schottky-type shot noise of a CQD interferometer, for which not only RDM diagonal elements 
contribute to the current, but also off-diagonal elements contribute as well, with stationary 
values and corresponding tunneling-rates which could possibly both be negative.

\section{CQD in a series configuration}

In this section, we discuss coherent resonant tunneling through a series-CQD in the case of 
infinite inter-dot Coulomb repulsion, $U=\infty$. In this case, the corresponding tunneling 
amplitudes and the interferential tunneling vanish, $\Gamma_{L22}=\Gamma_{R11}=0$, 
$\Gamma_{L12}=\Gamma_{R12}=0$, and all $\widetilde{\Gamma}=0$. We also choose $\delta=0$ 
($\varepsilon_1=\varepsilon_2=\varepsilon_d$). 

For comparison with our previous QREs in the case of weak inter-dot hopping,\cite{Dong} we 
re-express the resulting generic QREs Eqs~(\ref{qrer00})-(\ref{qrerab}) in the SR for the case of 
a series configuration ($\Gamma_{L\chi}^\pm=\Gamma_{L11\chi}^\pm$ and $\Gamma_{R\chi}^\pm = 
\Gamma_{R22\chi}^\pm$):
\begin{subequations}
\bn
\dot \rho_{00} &=& - (\widetilde \Gamma_{L}^+ + \widetilde \Gamma_{R}^+ ) \rho_{00} + \widetilde 
\Gamma_{L}^- \rho_{11} + \widetilde \Gamma_{R}^- \rho_{22} \cr
&& + \frac{1}{4} \sin \theta (\Gamma_{L\alpha}^- + \Gamma_{R\alpha}^- - \Gamma_{L\beta}^- - 
\Gamma_{R\beta}^-) (\rho_{12}+ \rho_{21}), \cr
&& \label{qrer00s} \\
\dot \rho_{11} &=& \widetilde \Gamma_{L}^+ \rho_{00} - \widetilde \Gamma_{L}^- \rho_{11} + 
i\Omega (\rho_{21}- \rho_{12}) \cr
&& - \frac{1}{4} \sin \theta (\Gamma_{L\alpha}^- - \Gamma_{L\beta}^-) (\rho_{12}+ \rho_{21}), 
\label{qrer11s} \\
\dot \rho_{22} &=& \widetilde \Gamma_{R}^+ \rho_{00} - \widetilde \Gamma_{R}^- \rho_{22} - 
i\Omega (\rho_{21}- \rho_{12}) \cr
&& - \frac{1}{4} \sin \theta (\Gamma_{R\alpha}^- - \Gamma_{R\beta}^-) (\rho_{12}+ \rho_{21}), 
\label{qrer22s}\\
\dot \rho_{12} &=& i 2\delta \rho_{12} - i\Omega (\rho_{11} - \rho_{22}) - \frac{1}{2} 
(\widetilde \Gamma_{L}^- + \widetilde \Gamma_{R}^-) \rho_{12} \cr
&& + \frac{1}{4} \sin \theta (\Gamma_{L\alpha}^+ + \Gamma_{R\alpha}^+ - \Gamma_{L\beta}^+ - 
\Gamma_{R\beta}^+) \rho_{00} \cr
&& -\frac{1}{4} \sin \theta (\Gamma_{R\alpha}^- - \Gamma_{R\beta}^-) \rho_{11} - \frac{1}{4} \sin 
\theta (\Gamma_{L\alpha}^- - \Gamma_{L\beta}^-) \rho_{22}, \cr
&& \label{qrer12s}
\en
with $\rho_{00}+ \rho_{11} + \rho_{22}=1$ and $\rho_{21}=\rho_{12}^*$, in which
\bq
\widetilde \Gamma_{L}^\pm = c_{\theta}^2 \Gamma_{L\alpha}^\pm + s_{\theta}^2 \Gamma_{L\beta}^\pm,
\eq
and
\bq
\widetilde \Gamma_{R}^\pm = s_{\theta}^2 \Gamma_{R\alpha}^\pm + c_{\theta}^2 \Gamma_{R\beta}^\pm,
\eq
\end{subequations}
denotes the effective tunneling rates between the first(second) dot and the left(right) lead. The 
current becomes
\bq
I_L = \widetilde\Gamma_L^+ \rho_{00} - \widetilde \Gamma_{L}^- \rho_{11} - \frac{1}{4} \sin 
\theta (\Gamma_{L\alpha}^- - \Gamma_{L\beta}^-) (\rho_{12} + \rho_{21}). \label{isL}
\eq

Comparing with our previous results [Eqs.~(39a)-(39c) of Ref.~\onlinecite{Dong}], we find some 
new features in our present QREs, Eqs.~(\ref{qrer00s})-(\ref{qrer12s}), in the SR: (1) The 
tunneling rates between dot and lead, $\widetilde \Gamma_{L/R}^{\pm}$, are modified to account 
for the effect of interdot hopping; (2) More interestingly, effective interdot hopping induced by 
tunneling to external electrodes provides a new contribution, which is dependent on the real part 
of the RDM off-diagonal elements, to the dynamics of the RDM diagonal elements stemming from 
quantum coherence [for instance, the fourth terms on the right hand side (rhs) of 
Eqs.~(\ref{qrer11s}) and (\ref{qrer22s})]; (3) Conversely, this effective interdot hopping also 
leads to a new damping of quantum superposition related to the diagonal elements themselves, as 
shown by the final three terms on the rhs of Eq.~(\ref{qrer12s}). Moreover, the real part of the 
off-diagonal elements make an explicit, direct contribution to the tunneling current and 
Schottky-type shot noise as well.

\begin{figure}[htb]
\includegraphics[height=6cm,width=8.5cm]{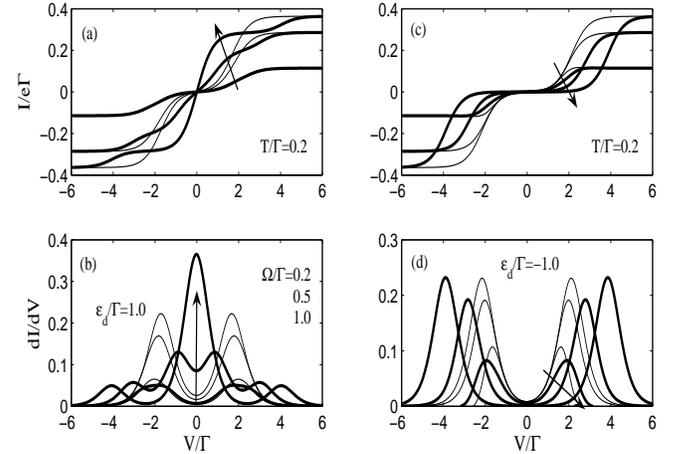}
\caption{Average current (a,c), and the differential conductance (b,d) vs. bias-voltage with 
several different interdot hopping strengths $\Omega/\Gamma=0.2$, $0.5$, and $1.0$. (a) and (b) 
are plotted for $\varepsilon_d/\Gamma=1.0$, (c) and (d) for $\varepsilon_d/\Gamma=-1.0$. The 
thick lines denote the results calculated by the present QREs, 
Eqs.~(\ref{qrer00s})-(\ref{qrer12s}) and Eq.~(\ref{isL}); while the thin lines arise from the 
previous QREs, Eqs.~(39a)-(39c) in Ref.~\onlinecite{Dong}. The temperature is fixed as 
$T/\Gamma=0.2$.}
\label{fig:comp1}
\end{figure}

In regard to the fact that our previous QREs are only valid for the case of weak interdot 
hopping, whereas our present QREs are derived without that limitation, it is interesting to note 
that the present generic QREs do reduce to the previous QREs in the limit $\Omega\rightarrow 0$ 
($\ll \Gamma_{L/R}$). In this situation, from Eq.~(\ref{eigenergy}) we have 
$\lambda_{\alpha(\beta)}=\varepsilon_d \pm \Omega$ ($\delta=0$), leading to $\lambda_{\alpha} 
\approx \lambda_{\beta}$ and thus $\Gamma_{\eta \alpha}^\pm \approx\Gamma_{\eta \beta}^\pm$. This 
approximate relation eliminates the additional quantum coherence terms of 
Eqs.~(\ref{qrer00s})-(\ref{qrer12s}) due to the tunneling-induced effective dot-dot hopping, and 
consequently the QREs reduce exactly to our previous results. For numerical verification of this 
assertion, we examine the effect of hopping ($\Omega/\Gamma=0.2\sim 1.0$) on resonant tunneling 
in Fig.~\ref{fig:comp1}, where we plot the measurable physical quantities, e.g. the occupation 
numbers in both dots, $\rho_{11}$ and $\rho_{22}$, the average current $I$, and the differential 
conductance $dI/dV$ as functions of bias-voltage for two cases, $\varepsilon_d/\Gamma=1.0$ and 
$-1.0$. Hereafter, we choose a completely symmetric geometry, $\Gamma_{L11}=\Gamma_{R22}=\Gamma$, 
and use $\Gamma$ as the unit of energy, and the current is normalized to $\frac{e}{\hbar} 
\Gamma$. We exhibit the calculated results using the present QREs, 
Eqs~(\ref{qrer11s})-(\ref{qrer12s}), by the thick lines and those of the previous QREs, 
Eqs.~(39a)-(39c) in Ref.~\onlinecite{Dong}, by the thin lines. 

\begin{figure}[htb]
\includegraphics[height=4.5cm,width=8cm]{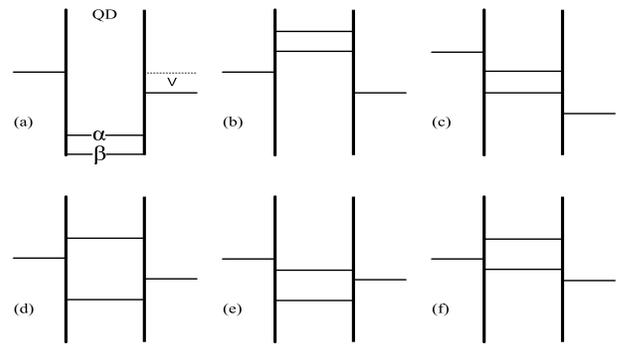}
\caption{Schematic energetic configurations for coherent tunneling through a CQD in the ER. The 
CQD behaves as a single QD with two levels $\alpha$ and $\beta$, both of which are coupled to 
both the left and right leads. In configurations (a)-(c), the additional quantum coherence terms 
automatically vanish in the QREs (\ref{qrer00s})-(\ref{qrer12s}) and in the current formula 
(\ref{isL}), while in (d)-(f), they play an important role in tunneling.}
\label{fig:energycon}
\end{figure}

It is obvious that there is no nontrivial difference in the separate calculations for these 
physical quantities in the case of weakest dot-dot hopping, $\Omega/\Gamma=0.2$. With increasing 
dot-dot hopping, the difference between the two theories becomes unambiguous and cannot be 
ignored. In particular, a zero-bias peak (ZBP) in the differential conductance, $dV/dI$, is found 
by the present theory for the system of $\varepsilon_d/\Gamma=1.0$ with dot-dot hopping 
$\Omega/\Gamma=1.0$, in contrast to the peak-splitting around zero-bias-voltage predicted by the 
previous QREs. On the other hand, for the system of $\varepsilon_d/\Gamma=-1.0$, both calculated 
results for differential conductance show a peak-splitting structure but they have different 
splitting widths. The transport properties are clearly asymmetric between the systems with 
$\varepsilon_d>0$ and $\varepsilon_d<0$, which have been already pointed out in linear transport 
regime.\cite{Konig} In fact, this asymmetric feature can be easily explained by examining the 
energetic structure of the present system. As mentioned above, the CQD in the ER can be mapped 
onto a model of a single QD with two levels, $\alpha$ and $\beta$, each connecting to both 
electrodes with different tunneling matrix elements. The physical picture is illustrated in 
Fig.~\ref{fig:energycon}, which includes a total of six configurations [(a)-(f)], showing the 
relative relations among the two eigen-energies and the Fermi energies of the two electrodes. For 
simplicity, we set $T=0$ in the following analysis.
In configurations (a) and (d), the CQD is always singly occupied by an electron, while it is 
always empty in configuration (b). The three cases have all vanishing first-order tunneling 
current and current noise because the two eigen-levels are both far away from resonance with the 
Fermi levels of the two electrodes.     
On the contrary, both eigen-levels are in resonance with the Fermi levels of the electrodes in 
configuration (c), which is actually equivalent to a picture of resonant tunneling under 
extremely large bias-voltage. It is therefore not unexpected that the resulting rates and current 
in this case are identical to the results obtained by Gurvitz and Prager for the same 
system,\cite{Gurvitz} as well as matching our previous results for the same 
conditions.\cite{Dong} 

Moreover, it is quite surprising that the current is also zero in the resonant configuration (e). 
We can ascribe this result to the strong Coulomb blockade effect, which blocks the entry of an 
additional electron into the singly-occupied CQD even though the upper eigen-level is in 
resonance with the Fermi levels of electrodes. The situation is different in configuration (f), 
in which only the lower eigen-level is in resonance with the Fermi levels of two leads and thus a 
nonzero current occurs. We note that this current is always smaller than the current in 
configuration (c) [This is only true for a series-CQD. While for a parallel-CQD, the situation 
may be opposite (see below)]. Actually, configuration (f) describes a picture of resonant 
tunneling under a small bias-voltage, in contrast to the case of large bias-voltage in 
configuration (c).  

\begin{figure}[htb]
\includegraphics[height=4cm,width=8.5cm]{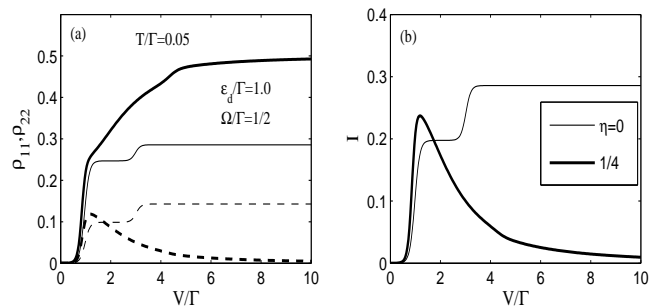}
\caption{(a) Occupation numbers $\rho_{11}$ (solid lines), $\rho_{22}$ (dashed lines) and (b)  
average current vs. bias-voltage with the bias-voltage-induced shifting factor $\eta=0$ (thin 
lines) and $1/4$ (thick lines) for $\varepsilon_d/\Gamma=1.0$, $\Omega/\Gamma=1/2$, and 
$T/\Gamma=0.05$.}
\label{fig:shift}
\end{figure}

It should be noted that our present QREs are also applicable for the cases with a detuning of the 
bare level energies, i.e., $\delta\neq 0$. Actually, spatial variation of bias-voltage on the 
series-CQD is quite a central issue in the present studies.\cite{Datta,Pati,Pedersen} Without 
loss of generality, we assume that the bare QD energies are shifted by the bias-voltage as 
$\varepsilon_{1(2)}=\varepsilon_d \pm \eta V$, and assume further that one fourth of the applied 
voltage drops at the contacts between the QDs and the leads, i.e., the bias-voltage-induced 
shifting factor $\eta=1/4$. As an illustration, we exhibit the effects of energy shifting on the 
electron occupation numbers of two QDs and the current in Fig.~\ref{fig:shift}. It is found that 
the main effect of the spatially varying electric field is to cause an appearance of negative 
differential conductance (NDC) in the current-voltage characteristics, which has been reported 
previously by nonequilibrium Green's function calculations\cite{Pati} in order to explain the 
experimental results in the Tour molecules.\cite{Tour} Here, we can give a simple interpretation 
for the appearance of NDC as follows: Due to the spatial variation of bias-voltage, the 
eigen-energies for the CQD become voltage-dependent, $\lambda_{\alpha(\beta)}=\varepsilon_d\pm 
\sqrt{\Omega^2+ \eta^2 V^2}$. This dependence remarkably modifies the electronic occupation 
behavior in comparison with no shifting case. Electrons have more opportunity to occupy the first 
QD for large bias-voltages, leading to decreasing of the occupation number of the second QD, 
$\rho_{22}$, with increasing bias-voltage as shown in Fig.~\ref{fig:shift}(a), which is reason of 
the appearance of NDC, because the nonequilibrium current is dominatively dependent on the 
occupation number of the second QD in the series configuration.

\section{CQD in a parallel configuration}

Focusing attention on the case of resonant tunneling through a parallel CQD, we again choose the 
optimal resonant condition, $\delta=0$, and we also assume the tunneling constants as 
$\Gamma_{L11}=\Gamma_{R22}= \Gamma$ and $\Gamma_{L22}=\Gamma_{R11}=\Gamma'$, thus 
$\Gamma_{L12}=\Gamma_{R12}=\sqrt{\Gamma \Gamma'}$, which yield
\begin{subequations}
\bn
\Gamma_{11\chi}^{\pm}&=& \Gamma f_L^\pm (\lambda_\chi) + \Gamma' f_{R}^\pm (\lambda_\chi), \\
\Gamma_{22\chi}^{\pm}&=& \Gamma' f_L^\pm (\lambda_\chi) + \Gamma f_{R}^\pm (\lambda_\chi), \\ 
\Gamma_{12\chi}^{\pm}&=& \sqrt{\Gamma \Gamma'} [ e^{i\varphi/2} f_L^\pm (\lambda_\chi) + 
e^{-i\varphi/2} f_{R}^\pm (\lambda_\chi)], \\
\Gamma_{21\chi}^{\pm}&=& \sqrt{\Gamma \Gamma'} [ e^{-i\varphi/2} f_L^\pm (\lambda_\chi) + 
e^{i\varphi/2} f_{R}^\pm (\lambda_\chi)].
\en
\end{subequations}
At first, we consider the case of $U\rightarrow\infty$. If dot-dot hopping vanishes, i.e., 
$\Omega=0$, the QREs coincide with our previous results in Ref.~\onlinecite{Ma}.  

\subsection{Simplified QREs in the SR for configurations (c) and (f)}\label{sec:parallelcandf}

The energetic configurations for coherent tunneling through a CQD interferometer are also 
depicted in Fig.~\ref{fig:energycon}.
Analogous to the series CQD, only configurations (c) and (f) bear nonzero tunneling current for 
the present parallel arrangement. 
For reference purposes, we therefore provide the simplified QREs written in the SR at zero 
temperature for these two configurations below.

For an extremely large bias-voltage [configuration (c)], these tunneling rates simplify further 
as $\Gamma_{11\alpha(\beta)}^+=\Gamma_{22\alpha(\beta)}^-=\Gamma$, $\Gamma_{22\alpha(\beta)}^+ = 
\Gamma_{11\alpha(\beta)}^-=\Gamma'$, $\Gamma_{12\alpha(\beta)}^+ = \Gamma_{21\alpha(\beta)}^- = 
\sqrt{\Gamma\Gamma'} e^{i\varphi/2}$, $\Gamma_{12\alpha(\beta)}^- = \Gamma_{21\alpha(\beta)}^+ = 
\sqrt{\Gamma\Gamma'} e^{-i\varphi/2}$. The simplified QREs in terms of the SR become finally
\begin{subequations}
\bn
\dot{\rho}_{00}&=& \Gamma' \rho_{11} + \Gamma \rho_{22} - (\Gamma + \Gamma') \rho_{00} \cr
&& + \sqrt{\Gamma \Gamma'} (e^{i\varphi/2} \rho_{12} + e^{-i\varphi/2} \rho_{21}), \label{rc0} \\
\dot{\rho}_{11}&=& \Gamma \rho_{00}  - \Gamma' \rho_{11} + i\Omega (\rho_{21}-\rho_{12}) \cr
&& - \frac{1}{2} \sqrt{\Gamma \Gamma'} (e^{i\varphi/2} \rho_{12} + e^{-i\varphi/2} \rho_{21}), 
\label{rc1} \\
\dot{\rho}_{22}&=& \Gamma' \rho_{00} - \Gamma \rho_{22} - i\Omega (\rho_{21}-\rho_{12}) \cr
&& - \frac{1}{2} \sqrt{\Gamma \Gamma'} (e^{i\varphi/2} \rho_{12} + e^{-i\varphi/2} \rho_{21}), 
\label{rc2} \\
\dot{\rho}_{12} &=& i 2 \delta \rho_{12} - i\Omega (\rho_{11}-\rho_{22}) - \frac{1}{2} ( \Gamma + 
\Gamma' ) \rho_{12} \cr
&& + \sqrt{\Gamma \Gamma'} e^{i\varphi/2} \rho_{00} - \frac{1}{2} \sqrt{\Gamma \Gamma'} 
e^{-i\varphi/2} (\rho_{11} + \rho_{22}), \cr
&& \label{rc4}
\en
\end{subequations}
along with the normalization relation, $\rho_{00}+ \rho_{11}+ \rho_{22}=1$, and 
$\rho_{21}=\rho_{12}^{*}$. As expected, we note that the resulting QREs, 
Eqs.~(\ref{rc0})-(\ref{rc4}), coincide with those of our previous derivation for the same system 
analyzed in Ref.~\onlinecite{Dong1}.  
The current becomes
\bn
I_L &=& (\Gamma + \Gamma') \rho_{00}, \label{currentLc} \\
I_R &=& -[\Gamma' \rho_{11} + \Gamma \rho_{22} + \sqrt{\Gamma\Gamma'} (e^{i \varphi/2} \rho_{12} 
+ e^{-i\varphi/2} \rho_{21})]. \nonumber
\en

Moreover, the system of configuration (f) yields  
\begin{subequations}
\begin{eqnarray}       
\dot\rho_{11} &=& (\Gamma s_\theta^2 - \frac{1}{2} \sqrt{\Gamma\Gamma'} \sin\theta 
\cos\frac{\varphi}{2})\rho_{00} + i \Omega (\rho_{21}-\rho_{12}) \cr
&& - (\Gamma'+\Gamma c_\theta^2 + \frac{1}{2} \sqrt{\Gamma\Gamma'} \sin\theta 
\cos\frac{\varphi}{2}) \rho_{11} \cr
&& - \frac{1}{2}\sqrt{\Gamma\Gamma'} (e^{i\varphi/2} \rho_{12}
+ e^{-i\varphi/2} \rho_{21}) \cr
&& - \frac{1}{2}\sqrt{\Gamma\Gamma'} s_\theta^2 (e^{-i\varphi/2} \rho_{12} + e^{i\varphi/2} 
\rho_{21}) \cr
&& - \frac{1}{4}\Gamma \sin\theta (\rho_{12}+\rho_{21}),
\end{eqnarray}
\begin{eqnarray}       
\dot\rho_{22} &=& (\Gamma' c_\theta^2 - \frac{1}{2} \sqrt{\Gamma\Gamma'} \sin\theta 
\cos\frac{\varphi}{2}) \rho_{00} - i\Omega(\rho_{21}-\rho_{12}) \cr
&& - (\Gamma+\Gamma' s_\theta^2 + \frac{1}{2} \sqrt{\Gamma\Gamma'} \sin\theta 
\cos\frac{\varphi}{2}) \rho_{22} \cr        
&& - \frac{1}{2} \sqrt{\Gamma\Gamma'} (e^{i\varphi/2} \rho_{12}+ e^{-i\varphi/2} \rho_{21}) \cr
&& -\frac{1}{2} \sqrt{\Gamma\Gamma'} c_\theta^2 (e^{-i\varphi/2} \rho_{12} + e^{i\varphi/2} 
\rho_{21}) \cr
&& - \frac{1}{4}\Gamma'\sin\theta(\rho_{12}+\rho_{21}),
\end{eqnarray}
\begin{eqnarray}   
\dot\rho_{12}&=& 2i\delta\rho_{12} - i\Omega(\rho_{11}-\rho_{22}) \cr
&& - \frac{1}{2}[ \sqrt{\Gamma\Gamma'} \sin\theta e^{i\varphi/2} 
+ \Gamma'(s_\theta^2+1) + \Gamma (c_\theta^2 + 1)] \rho_{12} \cr
&& + [\frac{1}{2} \sqrt{\Gamma\Gamma'} e^{i\varphi/2} - \frac{1}{4} (\Gamma+\Gamma') \sin\theta] 
\rho_{00} \cr
&& - \frac{1}{2} \sqrt{\Gamma\Gamma'} e^{-i\varphi/2} (\rho_{11}+\rho_{22}) \cr
&& -\frac{1}{2} \sqrt{\Gamma\Gamma'} c_\theta^2 e^{i\varphi/2} \rho_{11}
- \frac{1}{4} \Gamma'\sin\theta \rho_{11} \cr
&& - \frac{1}{2} \sqrt{\Gamma\Gamma'} s_\theta^2 e^{i\varphi/2} \rho_{22}
- \frac{1}{4} \Gamma\sin\theta \rho_{22}.
\end{eqnarray}
\end{subequations}
The simplified current expressed in the ER is given by
\bn
I_L &=& 2[c_{\theta}^2 \Gamma' + s_{\theta}^2 \Gamma - \sqrt{\Gamma\Gamma'} \sin \theta \cos 
(\varphi/2)] \rho_{00} \cr
&& - 2 [c_{\theta}^2 \Gamma + s_{\theta}^2 \Gamma' + \sqrt{\Gamma\Gamma'} \sin \theta \cos 
(\varphi/2)] \rho_{\alpha\alpha} \cr
&& \hspace{-1cm} - [\frac{1}{2} (\Gamma - \Gamma') \sin \theta + s_{\theta}^2 
\sqrt{\Gamma\Gamma'} e^{i\varphi/2} - c_{\theta}^2 \sqrt{\Gamma\Gamma'} e^{-i\varphi/2}] 
\rho_{\alpha\beta} \cr
&& \hspace{-1cm} - [\frac{1}{2} (\Gamma - \Gamma') \sin \theta - c_{\theta}^2 
\sqrt{\Gamma\Gamma'} e^{i\varphi/2} + s_{\theta}^2 \sqrt{\Gamma\Gamma'} e^{-i\varphi/2}] 
\rho_{\beta\alpha}, \cr
&& \label{currentLf}
\en
which is useful for the analysis of the AB interference effect in this configuration.

\subsection{Calculations and discussion: AB interference effect between two pathways}

\begin{figure}[htb]
\mbox{\includegraphics[height=8cm,width=4cm]{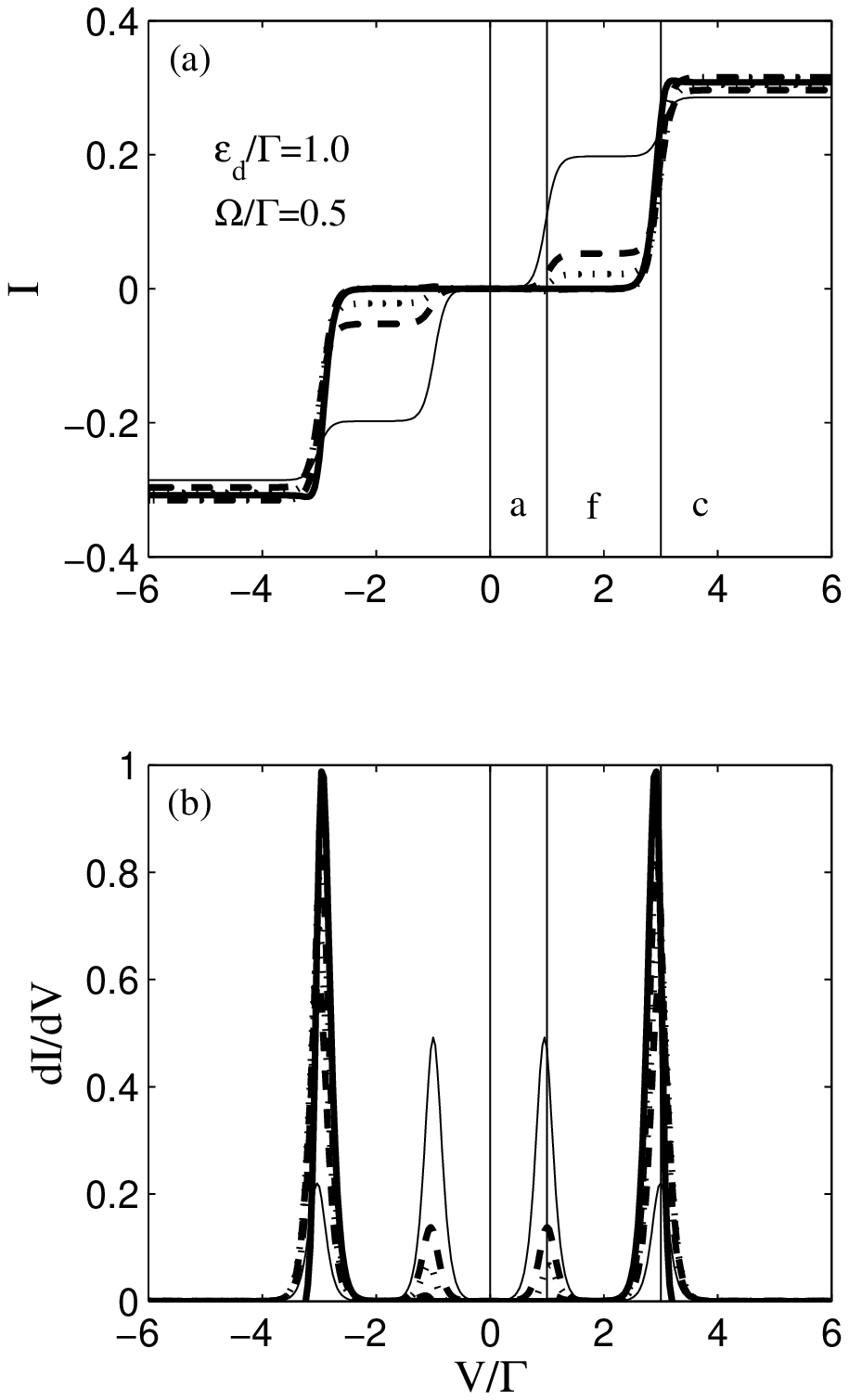}
\includegraphics[height=8cm,width=4cm]{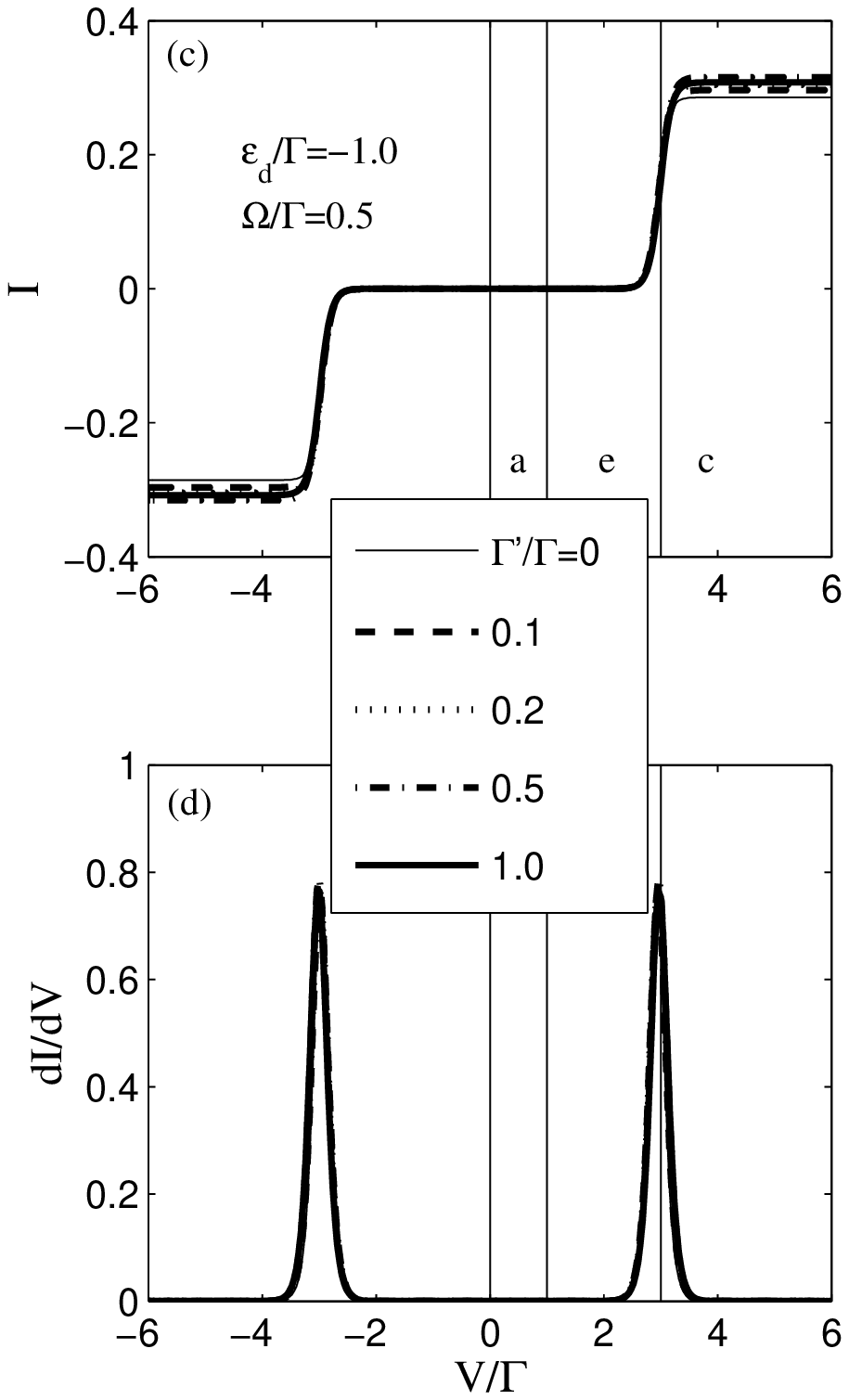}}
\caption{(a,c) the average current, and (b,d) the differential conductance $\frac{dI}{dV}$ vs. 
bias-voltage for several increasing tunnel-coupling strength values $\Gamma'/\Gamma=0$, $0.1$, 
$0.2$, $0.5$, and $1.0$ of the additional path without magnetic-flux, $\varphi=0$, for the 
systems with $\varepsilon_d/\Gamma=1.0$ (a,b) and $\varepsilon_d/\Gamma=-1.0$ (c,d). Other 
parameters are $\Omega/\Gamma=0.5$ and $T/\Gamma=0.05$.}
\label{fig:parallel1}
\end{figure}

Below, we present numerical analyses for first-order resonant tunneling through a CQD 
interferometer based on the QREs of Eqs.~(\ref{qrer00})-(\ref{qrerab}), and the current formula, 
Eq.~(\ref{currentL}). To start, we emphasize the interference effect on tunneling due to the 
additional pathway. As an illustration, we carry out calculations by changing the tunnel-coupling 
strength $\Gamma'$ of the additional path from $0$ (the series configuration) to $\Gamma$ (the 
completely symmetrical parallel configuration), as shown in Fig.~\ref{fig:parallel1} for two 
cases, $\varepsilon_d/\Gamma=1.0$ and $-1.0$, respectively, in the absence of magnetic-flux, 
$\varphi=0$.           

We note that the interference effect significantly changes transport properties when the system 
is in configuration (f) for $\varepsilon_d/\Gamma>0$, while it has little influence on tunneling 
when $\varepsilon_d/\Gamma<0$. This asymmetry of the interference pattern is also due to strong 
interaction. As displayed in Fig.~\ref{fig:parallel1}(c), the eigen-level $\beta$ is always 
singly-occupied in configurations (a) and (e), which blocks entry of an additional electron to 
the CQD, leading to vanishing current, irrespective of whether the additional tunnel-path is 
available or not. Increasing the tunnel rate $\Gamma'$ of the additional pathway suppresses the 
current until it entirely vanishes in the case of a completely symmetrical parallel geometry, 
$\Gamma'=\Gamma$, for the bias-voltage window $3>|V/\Gamma|>1$ of the system with 
$\varepsilon_d/\Gamma=1.0$ and $\Omega/\Gamma=1/2$ [configuration (f)] as depicted in 
Fig.~\ref{fig:parallel1}(a). Accordingly, the differential conductance finally develops a 
two-peak structure located at the bias-voltage values separating configurations (f) and (c) with 
an enhanced peak height [Fig.~\ref{fig:parallel1}(b)]. This behavior can be intuitively 
interpreted as a result of perfectly destructive quantum interference between the additional 
pathway and the original one for configuration (f).

\subsection{Finite-bias-induced AB oscillations and magnetic-flux-induced negative differential 
conductance}\label{sec:CQDparallelABoscillation}

In this subsection we examine the magnetic-flux dependence of coherent tunneling through the 
parallel CQD. Above, we found that nonzero tunneling current occurs only for configurations (f) 
(finite bias-voltage) and (c) (infinite bias-voltage). We consider these two cases.  

\begin{figure}[htb]
\includegraphics[height=7.cm,width=6cm]{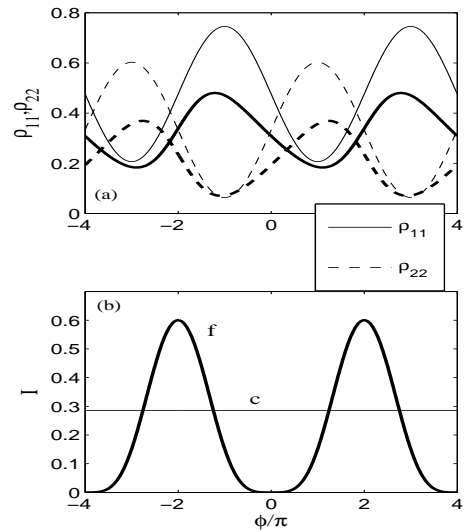}
\caption{Calculated magnetic-flux dependence of the dot occupations $\rho_{11}$ and $\rho_{22}$ 
(a) and the tunneling current (b) for a CQD with $\Gamma'/\Gamma=0.5$ and $\Omega/\Gamma=0.5$ in 
configurations (c) (thin lines) and (f) (thick lines), respectively.}
\label{fig:vphi}
\end{figure}

We determine the nonequilibrium dot occupations $\rho_{11(22)}$ and tunneling current $I$ based 
on the simplified QREs of Sec.~\ref{sec:parallelcandf}. The calculated results are illustrated in 
Fig.~\ref{fig:vphi} for a CQD with $\Gamma'/\Gamma=0.5$ and $\Omega/\Gamma=0.5$. Clearly, the dot 
occupations exhibit periodic oscillations with period $4\pi$ for both cases. Actually, our 
previous studies have already shown that interdot hopping yields periodicity of the AB 
oscillation as $4\pi$.\cite{Dong1} We further observe that $\rho_{11}\sim \sin(\varphi/2)$ and 
$\rho_{22} \sim -\sin (\varphi/2)$ for configuration (c), leading to a magnetic-flux independent 
result for $\rho_{00}=1-2(\rho_{11} + \rho_{22})$.
Since $I/\Gamma \sim \rho_{00}$ [Eq.~(\ref{currentLc})], the current of configuration (c) is 
consequently independent of magnetic-flux, as shown in Fig.~\ref{fig:vphi}(b). This result is 
quite surprising. In fact, if dissipative mechanisms are considered, $\rho_{22}$ may not be 
completely out of phase with $\rho_{11}$ (i.e., partial interference), thus the current of 
configuration (c) could become a periodic function of magnetic-flux.

\begin{figure}[htb]
\includegraphics[height=7.cm,width=6cm]{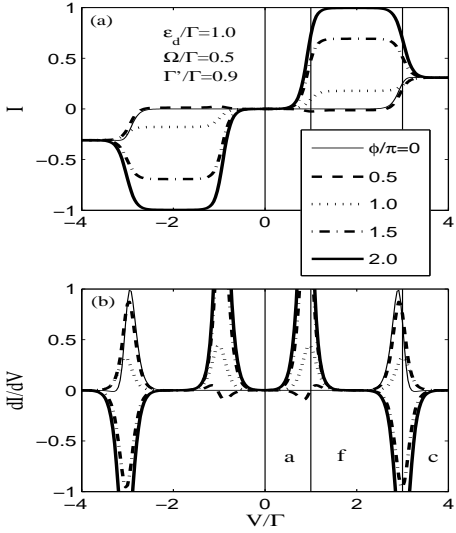}
\caption{Calculated tunneling current $I$ (a), and differential conductance $\frac{dI}{dV}$ (b), 
for a CQD with $\Gamma'/\Gamma=0.9$,  $\varepsilon_d/\Gamma=1.0$ and $\Omega/\Gamma=0.5$ for 
various magnetic-fluxes ($\varphi/\pi=0$, $0.5$, $1.0$, $1.5$, and $2.0$). The temperature is 
$T/\Gamma=0.05$.}
\label{fig:vv1}
\end{figure}

The current of configuration (f) is complex according to Eq.~(\ref{currentLf}). The results of 
the preceding subsection show that perfectly destructive quantum interference causes the current 
to vanish for zero magnetic-flux, $\varphi=0$, in this configuration 
[Fig.~\ref{fig:parallel1}(a)]. Application of a magnetic field will induce variation of the 
interference effect from a destructive pattern to a constructive one.      
Therefore, one can expect that the enclosed magnetic-flux, $\varphi=\pm 2\pi$, will enhance the 
current due to perfectly constructive interference. A striking result we obtain is that the 
enhanced current may be much larger than the constant current of configuration (c), indicating 
the appearance of a NDC at the boundary between configurations (f) and (c). The variation of the 
$I$-$V$ characteristic and differential conductance with varying magnetic-flux are shown in 
Fig.~\ref{fig:vv1}.           

\subsection{Finite interdot Coulomb repulsion $U$}

Here we examine the transport of a CQD in the case of weak dot-dot Coulomb interaction, 
$U=2\Gamma$.

\begin{figure}[htb]
\includegraphics[height=7cm,width=8cm]{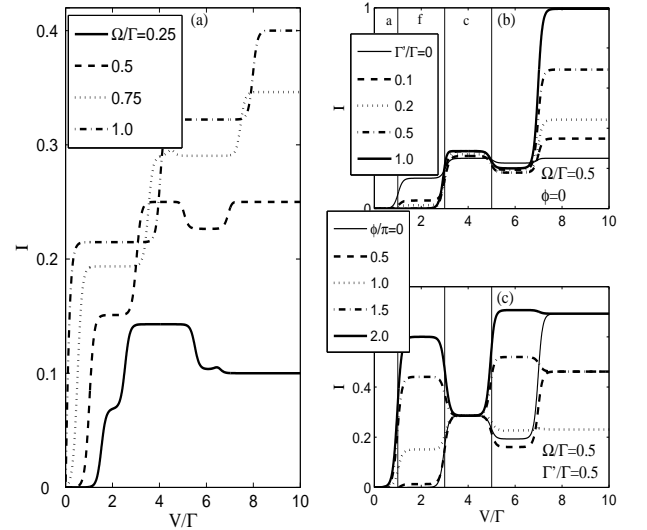}
\caption{Bias-dependent tunneling current $I$ for a series-CQD with various dot-dot hoppings (a), 
and for a parallel-CQD as functions of $\Gamma'$ (b) and magnetic-flux (c). The interdot Coulomb 
repulsion is set as $U=2\Gamma$, and other parameters are $\varepsilon_d/\Gamma=1.0$ and 
$T/\Gamma=0.05$.}
\label{fig:finiteu}
\end{figure}

Figure \ref{fig:finiteu}(a) plots the calculated current for a series-CQD with 
$\varepsilon_d/\Gamma=1.0$ and various interdot hoppings, $\Omega/\Gamma=0.25\sim 1.0$. Different 
from the case of infinite $U$, a NDC is observed around $2(\lambda_\beta+U)$ even for a 
series-CQD with weak interdot coupling $\Omega/\Gamma\leq 0.5$. This NDC behavior was first 
reported in our previous study\cite{Dong} and was ascribed to decoherence effect in coherent 
tunneling due to the coupling to leads.\cite{Djuric2,Pedersen} From Fig.~\ref{fig:finiteu}(b) and 
(c), we observe that this NDC is robust against the additional pathway $\Gamma'$ in the absence 
of magnetic-flux and it is tuned to positive differential conductance (PDC) when applying the 
magnetic field. It is worth noting that in the case of weak interdot Coulomb repulsion, above 
mentioned magneticc-flux-controlled NDC still remain around the boundary between configurations 
(f) and (c) [here configuration (c) is at the bias voltage region $3<V<5$].

\section{Zero-frequency shot noise}

\subsection{Two-terminal number-resolved QREs and MacDonald's formula}

This section is concerned with the discussion of zero-frequency current noise of the CQD 
interferometer in the case of $U\rightarrow\infty$. For this purpose, we employ MacDonald's 
formula for shot noise\cite{MacDonald} based on a number-resolved version of the QREs describing 
the number of completed tunneling events.\cite{Chen} This can be straightforwardly derived from 
the established QREs, Eqs.~(\ref{qrer00})--(\ref{qrerab}). We introduce the two-terminal 
number-resolved density matrices $\rho_{\chi\chi'}^{(n,m)}(t)$, representing the probability that 
the system is in the electronic state $|\chi\rangle$ (for $\chi=\chi'$) or in the quantum 
superposition state (for $\chi\neq \chi'$) at time $t$ together with $n(m)$ electrons occupying 
the left(right) lead due to tunneling events. Obviously, $\rho_{\chi\chi'}(t)=\sum_{n,m} 
\rho_{\chi\chi'}^{(n,m)}(t)$ and the resulting two-terminal
number-resolved QREs for arbitrary bias-voltage and interdot hopping are:
\begin{subequations}
\bn
\dot \rho_{00}^{(n,m)} &=& - [ c_{\theta}^2 (\Gamma_{11\alpha}^+ + \Gamma_{22\beta}^+) + 
s_{\theta}^2 ( \Gamma_{11\beta}^+ + \Gamma_{22\alpha}^+) \cr
&& +\frac{1}{2} \sin \theta (\Gamma_{12\alpha}^+ + \Gamma_{21\alpha}^+ - \Gamma_{12\beta}^+ 
-\Gamma_{21\beta}^+) ] \rho_{00}^{(n,m)} \cr
&& \hspace{-1.5cm} + [ c_{\theta}^2 \Gamma_{L11\alpha}^- + s_{\theta}^2 \Gamma_{L22\alpha}^- + 
\frac{1}{2} \sin \theta ( \Gamma_{L12\alpha}^- + \Gamma_{L21\alpha}^-)] 
\rho_{\alpha\alpha}^{(n-1,m)} \cr
&& \hspace{-1.5cm} + [ c_{\theta}^2 \Gamma_{R11\alpha}^- + s_{\theta}^2 \Gamma_{R22\alpha}^- + 
\frac{1}{2} \sin \theta ( \Gamma_{R12\alpha}^- + \Gamma_{R21\alpha}^-)] 
\rho_{\alpha\alpha}^{(n,m-1)} \cr
&& \hspace{-1.5cm} + [ s_{\theta}^2 \Gamma_{L11\beta}^- + c_{\theta}^2 \Gamma_{L22\beta}^- - 
\frac{1}{2} \sin \theta (\Gamma_{L12\beta}^- + \Gamma_{L21\beta}^-)] \rho_{\beta\beta}^{(n-1,m)} 
\cr
&& \hspace{-1.5cm} + [ s_{\theta}^2 \Gamma_{R11\beta}^- + c_{\theta}^2 \Gamma_{R22\beta}^- - 
\frac{1}{2} \sin \theta (\Gamma_{R12\beta}^- + \Gamma_{R21\beta}^-)] \rho_{\beta\beta}^{(n,m-1)} 
\cr
&& \hspace{-1.5cm} + \frac{1}{4} \sin \theta (\Gamma_{L11\alpha}^- + \Gamma_{L11\beta}^- - 
\Gamma_{L22\alpha}^- - \Gamma_{L22\beta}^-) \cr
&& \times (\rho_{\alpha\beta}^{(n-1,m)} + \rho_{\beta\alpha}^{(n-1,m)}) \cr
&& \hspace{-1.5cm} + \frac{1}{4} \sin \theta (\Gamma_{R11\alpha}^- + \Gamma_{R11\beta}^- - 
\Gamma_{R22\alpha}^- - \Gamma_{R22\beta}^-) \cr
&& \times (\rho_{\alpha\beta}^{(n,m-1)} + \rho_{\beta\alpha}^{(n,m-1)}) \cr
&& \hspace{-1.5cm} + \frac{1}{2}[s_{\theta}^2 (\Gamma_{L12\alpha}^- + \Gamma_{L12\beta}^-) - 
c_{\theta}^2 (\Gamma_{L21\alpha}^- + \Gamma_{L21\beta}^-)] \rho_{\alpha\beta}^{(n-1,m)} \cr  
&& \hspace{-1.5cm} + \frac{1}{2}[s_{\theta}^2 (\Gamma_{R12\alpha}^- + \Gamma_{R12\beta}^-) - 
c_{\theta}^2 (\Gamma_{R21\alpha}^- + \Gamma_{R21\beta}^-)] \rho_{\alpha\beta}^{(n,m-1)} \cr  
&& \hspace{-1.5cm} - \frac{1}{2}[c_{\theta}^2 (\Gamma_{L12\alpha}^- + \Gamma_{L12\beta}^-) - 
s_{\theta}^2 (\Gamma_{L21\alpha}^- + \Gamma_{L21\beta}^-)] \rho_{\beta\alpha}^{(n-1,m)} \cr 
&& \hspace{-1.5cm} - \frac{1}{2}[c_{\theta}^2 (\Gamma_{R12\alpha}^- + \Gamma_{R12\beta}^-) - 
s_{\theta}^2 (\Gamma_{R21\alpha}^- + \Gamma_{R21\beta}^-)] \rho_{\beta\alpha}^{(n,m-1)}, \cr
&& \label{qrer00nr} 
\en
\bn
\dot \rho_{\alpha\alpha}^{(n,m)} &=& [c_{\theta}^2 \Gamma_{L11\alpha}^+ + s_{\theta}^2 
\Gamma_{L22\alpha}^+ + \frac{1}{2} \sin \theta ( \Gamma_{L12\alpha}^+ + \Gamma_{L21\alpha}^+)] 
\cr
&& \times \rho_{00}^{(n+1,m)} \cr
&& \hspace{-1cm} + [c_{\theta}^2 \Gamma_{R11\alpha}^+ + s_{\theta}^2 \Gamma_{R22\alpha}^+ + 
\frac{1}{2} \sin \theta ( \Gamma_{R12\alpha}^+ + \Gamma_{R21\alpha}^+)] \cr
&& \times \rho_{00}^{(n,m+1)} \cr
&& \hspace{-1cm} - [c_{\theta}^2 \Gamma_{11\alpha}^- + s_{\theta}^2 \Gamma_{22\alpha}^- + 
\frac{1}{2} \sin \theta (\Gamma_{12\alpha}^- + \Gamma_{21\alpha}^-)] \rho_{\alpha\alpha}^{(n,m)} 
\cr
&& \hspace{-1cm} - \frac{1}{4} \sin \theta (\Gamma_{11\beta}^- - \Gamma_{22\beta}^-) 
(\rho_{\alpha\beta}^{(n,m)} + \rho_{\beta\alpha}^{(n,m)}) \cr
&& \hspace{-1cm} - \frac{1}{2} (s_{\theta}^2 \Gamma_{12\beta}^- - c_{\theta}^2 
\Gamma_{21\beta}^-) \rho_{\alpha\beta}^{(n,m)} \cr
&& \hspace{-1cm} - \frac{1}{2} (s_{\theta}^2 \Gamma_{21\beta}^- - c_{\theta}^2 \Gamma_{12\beta}^- 
) \rho_{\beta\alpha}^{(n,m)} , \label{qreraanr}
\en
\bn
\dot \rho_{\beta\beta}^{(n,m)} &=& [s_{\theta}^2 \Gamma_{L11\beta}^+ + c_{\theta}^2 
\Gamma_{L22\beta}^+ - \frac{1}{2} \sin \theta ( \Gamma_{L12\beta}^+ + \Gamma_{L21\beta}^+)] \cr
&& \times \rho_{00}^{(n+1,m)} \cr
&& \hspace{-1cm} +[s_{\theta}^2 \Gamma_{R11\beta}^+ + c_{\theta}^2 \Gamma_{R22\beta}^+ - 
\frac{1}{2} \sin \theta ( \Gamma_{R12\beta}^+ + \Gamma_{R21\beta}^+)] \cr
&& \times \rho_{00}^{(n,m+1)} \cr
&& \hspace{-1cm} - [s_{\theta}^2 \Gamma_{11\beta}^- + c_{\theta}^2 \Gamma_{22\beta}^- - 
\frac{1}{2} \sin \theta (\Gamma_{12\beta}^- + \Gamma_{21\beta}^-)] \rho_{\beta\beta}^{(n,m)} \cr
&& \hspace{-1cm} - \frac{1}{4} \sin \theta (\Gamma_{11\alpha}^- - \Gamma_{22\alpha}^-) 
(\rho_{\alpha\beta}^{(n,m)} + \rho_{\beta\alpha}^{(n,m)}) \cr
&& \hspace{-1cm} - \frac{1}{2} (s_{\theta}^2 \Gamma_{12\alpha}^- - c_{\theta}^2 
\Gamma_{21\alpha}^-) \rho_{\alpha\beta}^{(n,m)} \cr
&& \hspace{-1cm} - \frac{1}{2} (s_{\theta}^2 \Gamma_{21\alpha}^- - c_{\theta}^2 
\Gamma_{12\alpha}^- ) \rho_{\beta\alpha}^{(n,m)} ,
\en
\bn
\dot \rho_{\alpha\beta}^{(n,m)} &=& i2\Delta \rho_{\alpha\beta}^{(n,m)} \cr
&& \hspace{-1.2cm} + [ \frac{1}{4} \sin \theta (\Gamma_{L11\alpha}^+ + \Gamma_{L11\beta}^+ - 
\Gamma_{L22\alpha}^+ - \Gamma_{L22\beta}^+) \cr
&& \hspace{-1.2cm} - \frac{1}{2} c_{\theta}^2 (\Gamma_{L12\alpha}^+ + \Gamma_{L12\beta}^+) + 
\frac{1}{2} s_{\theta}^2 (\Gamma_{L21\alpha}^+ + \Gamma_{L21\beta}^+)] \rho_{00}^{(n+1,m)} \cr
&& \hspace{-1.2cm} + [ \frac{1}{4} \sin \theta (\Gamma_{R11\alpha}^+ + \Gamma_{R11\beta}^+ - 
\Gamma_{R22\alpha}^+ - \Gamma_{R22\beta}^+) \cr
&& \hspace{-1.2cm} - \frac{1}{2} c_{\theta}^2 (\Gamma_{R12\alpha}^+ + \Gamma_{R12\beta}^+) + 
\frac{1}{2} s_{\theta}^2 (\Gamma_{R21\alpha}^+ + \Gamma_{R21\beta}^+)] \rho_{00}^{(n,m+1)} \cr
&& \hspace{-1.2cm} -\frac{1}{2} [ c_{\theta}^2 (\Gamma_{11\alpha}^- + \Gamma_{22\beta}^-) + 
s_{\theta}^2 (\Gamma_{11\beta}^- + \Gamma_{22\alpha}^-) \cr
&& \hspace{-1.2cm} - \frac{1}{2} \sin \theta (\Gamma_{12\alpha}^+ + \Gamma_{21\alpha}^+ - 
\Gamma_{12\beta}^+ - \Gamma_{21\beta}^+)] \rho_{\alpha\beta}^{(n,m)} \cr
&& \hspace{-1.2cm} + \frac{1}{2} [ c_{\theta}^2 \Gamma_{12\alpha}^- - s_{\theta}^2 
\Gamma_{21\alpha}^- - \frac{1}{2} \sin \theta (\Gamma_{11\alpha}^- - \Gamma_{22\alpha}^-)] 
\rho_{\alpha\alpha}^{(n,m)} \cr
&& \hspace{-1.2cm} + \frac{1}{2} [ c_{\theta}^2 \Gamma_{12\beta}^- - s_{\theta}^2 
\Gamma_{21\beta}^- - \frac{1}{2} \sin \theta (\Gamma_{11\beta}^- - \Gamma_{22\beta}^-)] 
\rho_{\beta\beta}^{(n,m)}.\cr
&& \label{qrerabnr}
\en
\end{subequations}
  
The current of lead $\eta$ can be evaluated as
\bq
I_{\eta} = \dot N_{\eta}(t) =\frac{d}{dt} \sum_{n,m} n_{\eta} P^{(n,m)}(t) {\Big 
|}_{t\rightarrow\infty}, \label{Inr}
\eq
where
\bq
P^{(n,m)}(t) = \rho_{00}^{(n,m)}(t)+ \rho_{11}^{(n,m)}(t)+ \rho_{22}^{(n,m)}(t) 
\eq
is the total probability of transferring $n(m)$ electrons into the left(right) lead by time $t$ 
and $n_{\eta}=n(m)$ if $\eta=L(R)$. It is easily verified that the current obtained from 
Eq.~(\ref{Inr}) by means of the number-resolved QREs, Eqs.~(\ref{qrer00nr})--(\ref{qrerabnr}), is 
exactly the same as that obtained from Eq.~(\ref{currentL}). The zero-frequency shot noise with 
respect to lead $\eta$ is similarly defined in terms of $P^{(n,m)}(t)$ as 
well:\cite{Elattari,Dong4,MacDonald,Chen}
\bq
S_{\eta}(0)=2\frac{d}{dt} \left [ \sum_{n,m} n_{\eta}^2 P^{(n,m)}(t) - (t I_{\eta})^2 \right ] 
{\Big |}_{t\rightarrow\infty}. \label{snnr}
\eq

To evaluate $S_{\eta}(0)$, we define an auxiliary function $G_{\chi\chi'}^{\eta}(t)$ as
\bq
G_{\chi\chi'}^{\eta}(t)=\sum_{n,m} n_{\eta} \rho_{\chi\chi'}^{(n,m)}(t),
\eq
whose equations of motion can be readily deduced employing the number-resolved QREs, 
Eqs~(\ref{qrer00nr})--(\ref{qrerabnr}), in matrix form: $\dot{\bm G}^{\eta}(t)={\cal 
M}_{\eta}{\bm G}^{\eta}(t) + {\cal G}_{\eta}{\bm \rho}(t)$ with ${\bm 
G}^{\eta}(t)=(G_{00}^{\eta},G_{\alpha\alpha}^{\eta},G_{\beta\beta}^{\eta}, 
G_{\alpha\beta}^{\eta},G_{\beta\alpha}^{\eta})^{T}$ and ${\bm 
\rho}(t)=(\rho_{00},\rho_{\alpha\alpha},\rho_{\beta\beta}, 
\rho_{\alpha\beta},\rho_{\beta\alpha})^{T}$. ${\cal M}_{\eta}$ and ${\cal G}_{\eta}$ can be read 
easily from Eqs.~(\ref{qrer00nr})--(\ref{qrerabnr}). 
Applying the Laplace transform to these equations yields
\bq
{\bm G}^{\eta}(s) = (s {\bm I}-{\cal M}_{\eta})^{-1} {\cal G}_{\eta} {\bm \rho}(s),
\eq
where ${\bm \rho}(s)$ is readily obtained by applying the Laplace transform to its equations of 
motion with the initial condition ${\bm \rho}(0)={\bm \rho}_{st}$ [${\bm \rho}_{st}$ denotes the 
stationary solution of the QREs, Eqs~(\ref{qrer00})--(\ref{qrerab})]. Due to the inherent 
long-time stability of the physical system under consideration, all real parts of nonzero poles 
of ${\bm \rho}(s)$ and ${\bm G}^{\eta}(s)$ are negative definite. Consequently,  the divergent 
terms arising in the partial fraction expansions of ${\bm \rho}(s)$ and ${\bm G}^{\eta}(s)$ as 
$s\rightarrow 0$ entirely determine the large-$t$ behavior of the auxiliary functions, i.e. the 
zero-frequency shot noise, Eq.~(\ref{snnr}). 

It is worth mentioning that (1) our two-terminal number-resolved QREs, 
Eqs.~(\ref{qrer00nr})--(\ref{qrerabnr}), facilitate evaluation of the bias-voltage dependent 
zero-frequency shot noise for arbitrary interdot hopping; (2) our calculations yield 
$S_{L}(0)=S_{R}(0)$.    

\subsection{Results and discussion}

We confined our studies to the case $\varepsilon_d>0$. The Fano factor, $F=S(0)/2I$, is used as 
the main tool to classify current.\cite{Blanter} It should be noted that because we use the 
formula $F=S(0)/2I$ to calculate it numerically, the Fano factor shown in the figures below is 
physically meaningful only when current increases above zero to avoid numerical divergence. Thus, 
our calculated Fano factor is physically meaningful only for configurations (f) and (c).    

\begin{figure}[htb]
\includegraphics[height=3.5cm,width=8cm]{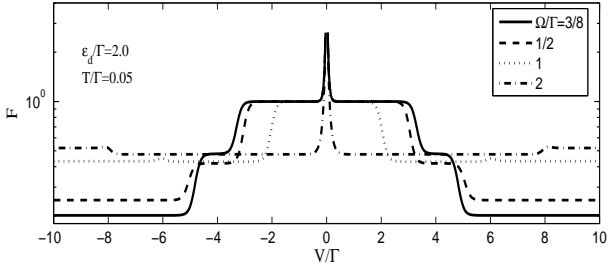}
\caption{Calculated Fano factor, $F=S(0)/2I$, for the series-connected CQD with 
$\varepsilon_d/\Gamma=2.0$ and various values of $\Omega/\Gamma$ ($3/8$, $1/2$, $1$, and $2$). 
The temperature is $T/\Gamma=0.05$.}
\label{fig:sn-series}
\end{figure}

To start, we discuss shot noise for a series-connected CQD with symmetric geometry, 
$\Gamma_{L11}=\Gamma_{R22}=\Gamma$. In the case of configuration (c), i.e., the extremely large 
bias-voltage limit with zero temperature, we obtain an analytical expression for the Fano 
factor:\cite{Dong4} 
\bq
F_c= \frac{80 x^4 - 8x^2 +1}{(1+12 x^2)^2}, \label{sn0-c}
\eq
with $x=\Omega/\Gamma$.
For configuration (f), i.e., with appropriately small bias-voltage, we have
\bq
F_f= \frac{8192 x^4 + 896 x^2 +441}{(21+128 x^2)^2}.
\eq
The results indicate sub-Poissonian shot noise, i.e. $F< 1$, in the effective bias-voltage regime 
for any dot-dot hopping strength, as shown in Fig.~\ref{fig:sn-series}. If the dot-dot hopping 
strength is set moderately weak, $\Omega/\Gamma\leq  0.978$ (which can be achieved by tuning the 
coupling potential between the two dots via applied gate voltage), we have $F_f>F_c$; otherwise, 
$F_f<F_c$. 

\begin{figure}[htb]
\includegraphics[height=3.5cm,width=8cm]{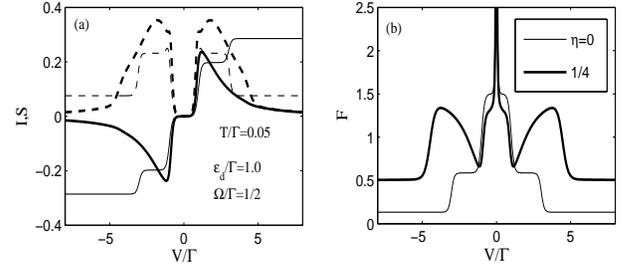}
\caption{(a) Zero-frequency shot noise $S(0)$ (dashed lines) and tunneling current $I$ (solid 
lines), and (b) the Fano factor $F$, for the series-CQD with $\varepsilon_d/\Gamma=1.0$, 
$\Omega/\Gamma=1/2$ and shifting factors $\eta=0$ and $1/4$. The temperature is $T/\Gamma=0.05$.}
\label{fig:sn-series-shift}
\end{figure}

We also examine the effect of the bias-voltage-induced shifting of bare levels on the shot noise 
in Fig.~\ref{fig:sn-series-shift}. It is shown that this shifting results in (i) a weak 
enhancement of the shot noise, i.e., a super-Poissonian noise to be a companion to the NDC when 
the eigen-level $\beta$ dominates in transport, and (ii) a constant Fano factor $F=1/2$ at the 
extremely large bias-voltage region.

\begin{figure}[htb]
\includegraphics[height=6.cm,width=8cm]{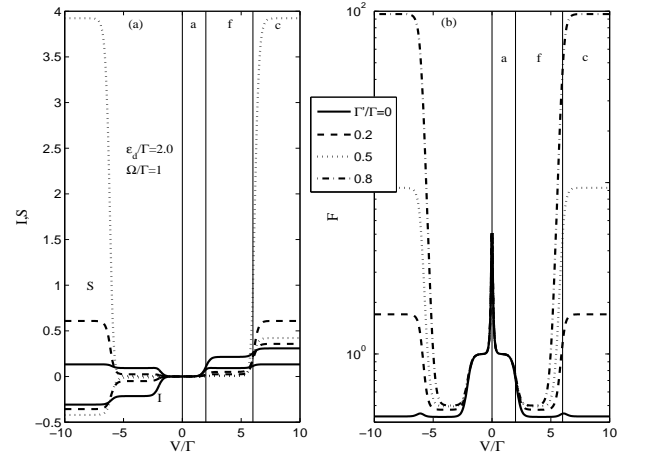}
\caption{(a) Zero-frequency shot noise $S(0)$ and tunneling current $I$, and (b) the Fano factor 
$F$, for the case $\varepsilon_d/\Gamma=2.0$ and $\Omega/\Gamma=1$ with various values of 
$\Gamma'/\Gamma$ in the absence of magnetic-flux. The temperature is $T/\Gamma=0.1$.}
\label{fig:sn-parallel}
\end{figure}

Focusing attention on the more general parallel CQD geometry, we discuss the quantum interference 
effect of the additional pathway on shot noise. 
Our calculated results in the absence of magnetic-flux are plotted in 
Figs.~\ref{fig:sn-parallel}, including the shot noise $S(0)$ (normalized to $\frac{e^2}{\hbar} 
\Gamma$), the current $I$ (normalized to $\frac{e}{\hbar} \Gamma$) for comparison, and the Fano 
factor. We find that the system exhibits a huge Fano factor with increasing tunneling rate of the 
additional branch in the case of configuration (c).\cite{Dong4} At zero temperature, we arrive at 
analytical expressions
\bn
I_c &=&\frac{4x^2 \Gamma (\gamma+1)} {(\gamma+1)^2+12 x^2}, \\
F_c &=& [(80\gamma^2+352\gamma+80)x^4 \cr
&& +(-8\gamma^4+160\gamma^3+336 \gamma^2+160\gamma-8)x^2 \cr 
&& +\gamma^6+10\gamma^5+31\gamma^4+44\gamma^3+31\gamma^2+10\gamma+1] \cr
&& \times (\gamma-1)^{-2} [(\gamma+1)^2+12 x^2]^{-2}, \label{sn0-c-phi0}
\en
with $x=\Omega/\Gamma$ and $\gamma=\Gamma'/\Gamma$. It is obvious that (1) in the case of a 
series-CQD, Eq.~(\ref{sn0-c-phi0}) reduces exactly to Eq.~(\ref{sn0-c}), exhibiting 
sub-Poissonian behavior; and in contrast, (2) $F_c>1$ with increasing $\gamma$.\cite{Dong4} 

It should be noted that all calculations in the present paper are performed under the assumption 
of full interference between the two pathways and infinite inter-dot Coulomb repulsion as well. 
If decoherence is taken into account due to some dissipative mechanisms, the degree of 
interference will naturally lessen, thus leading to a great suppression of the Fano factor. In 
particular, in the case of full noninterference, the Fano factor reduces to a constant $F=10/27$, 
exhibiting sub-Poissonian shot noise. On the other hand, it has been reported that the noise is 
always sub-Poissonian in the case of no inter-dot Coulomb interaction.\cite{Dong4}    

\begin{figure}[htb]
\includegraphics[height=7.cm,width=8cm]{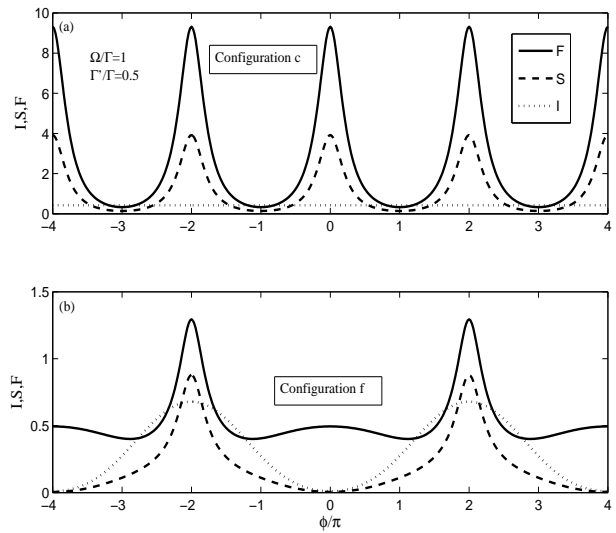}
\caption{Calculated magnetic-flux dependence of the zero-frequency shot noise $S(0)$, tunneling 
current $I$, and the Fano factor $F=S(0)/2I$ for a system with $\Omega/\Gamma=1.0$ and 
$\Gamma'/\Gamma=0.5$ in configuration (c) [panel (a)], and (f) [panel (b)] at zero temperature.}
\label{fig:snop}
\end{figure}

The phase effect on shot noise in a CQD interferometer tuned by magnetic-flux is of special 
interest for configurations (c) and (f). We exhibit the magnetic-flux dependence of current, shot 
noise, and Fano facor in Fig.~\ref{fig:snop} with $\Gamma'/\Gamma=0.5$ at zero temperature.   
Due to interdot hopping, both the current and shot noise exhibit periodic oscillations with 
period $4\pi$ (as well as the Fano factor) for configuration (f). The current nearly vanishes 
around zero magnetic-flux, $\varphi=0$, as indicated in Fig.~\ref{fig:vphi}(b) in 
Sec.~\ref{sec:CQDparallelABoscillation}; while the shot noise is suppressed much more due to the 
effect of perfectly destructive quantum interference. In contrast, constructive quantum 
interference enhances the zero-frequency shot noise more than the current, giving rise to 
super-Poissonian noise $F\simeq 5/4$ at $\varphi=\pm 2\pi$ even for configuration (f). On the 
other hand, the shot noise for configuration (c) displays different magnetic-flux dependence from 
that of configuration (f). Both $S(0)$ and $F$ are observed to behave as $\sim |\sin (\varphi)|$. 
Interestingly, varying magnetic-flux could change the shot noise from super-Poissonian, $F\simeq 
10$ at $\varphi=2n\pi$ ($n$ is an integer), to sub-Poissonian, $F\simeq 0.33$ at 
$\varphi=2(n+1)\pi$. Actually, we have derived an analytical expression for noise in 
configuration (c) and $\varphi=\pm \pi$:
\bn
F_c &=& [80 x^4- 8(\gamma^2+2\gamma+1)x^2+\gamma^4 \cr
&& +4\gamma^3+6\gamma^2+4\gamma+1] [(\gamma+1)^2+12 x^2]^{-2},
\en
indicating sub-Poissonian noise.   
It is also of interest to point out that at these values of magnetic-flux [$\varphi$ around 
$2(n+1)\pi$], the quantum interference effect induces NDC even for a very weak tunneling rate of 
the additional pathway, $\Gamma'/\Gamma=0.1$, as indicated in Figs.~\ref{fig:vphi} and 
\ref{fig:vv1} in Sec.~\ref{sec:CQDparallelABoscillation} [roughly $\varphi\geq 1.2\pi$], but 
super-Poissonian noise is not necessarily an accompaniment of this magnetic-flux-tuned NDC.

\section{Conclusions}

In summary, we have analyzed first-order transport
through a CQD AB interferometer with finite dot-dot hopping, determining the $I$-$V$ 
characteristics, zero-frequency current noise, and their magnetic-flux dependence. To accomplish 
this, we established generic QREs in terms of the ER employing a quantum Langevin equation 
approach in the weak tunneling limit. These QREs are valid for arbitrary temperature and 
bias-voltage, as well as arbitrary dot-dot hopping, improving upon our previous derivation which 
was limited by the restriction that inter-dot hopping be much weaker than dot-lead coupling, 
i.e., $\Omega \ll \Gamma$. 

We have also derived the current and Schottky-type shot noise formulae in terms of the RDM 
elements. Our theory proves that the previous scheme for evaluating the frequency-independent 
part of the ``classical" intrinsic shot noise (i.e., all tunneling processes providing 
contributions to current always yield positive contributions to the Schottky-type shot noise) 
remains valid in the formulation of quantum-based rate equations in the ER.

Employing the QREs derived here, we have systematically analyzed coherent resonant tunneling 
through a CQD in a series and a parallel configuration, respectively. By discussing variation of 
the energetic configuration with increasing bias voltage, we have explained the asymmetric 
transport property in the series CQD. We have also examined the effect of the 
bias-voltage-induced shifting of bare levels of the CQD and found the appearance of a NDC. For 
the parallel CQD, our numerical results have shown that (1) the current of configuration (c) is 
independent of magnetic-flux due to the combination of full interference and the strong Coulomb 
blockade effect; (2) AB oscillations emerge in the current in the small bias-voltage regime, 
i.e., the case of  configuration (f); (3) the current nearly vanishes completely around 
$\varphi=0$ due to perfect destructive interference, while it is greatly enhanced around 
$\varphi=2\pi$ due to constructive interference, and it may even be much larger than the current 
of configuration (c), suggesting the possibility of magnetic-flux-controllable NDC.  
  
Finally, we have investigated zero-frequency shot noise using MacDonald's formula by rewriting 
the fully developed QREs in terms of a two-terminal number-resolved density matrix form. The main 
result we obtained is that the combined effect of interference between two path branches and the 
infinite inter-dot Coulomb interaction may induce a huge Fano factor, which can also be 
controlled by manipulating magnetic-flux.

\begin{acknowledgments} 

This work was supported by Projects of the National Science Foundation of China, the Shanghai 
Municipal Commission of Science and Technology, the Shanghai Pujiang Program, and Program for New 
Century Excellent Talents in University (NCET). NJMH was supported by the DURINT program 
administered by the US Army Research Office, DAAD Grant
No.19-01-1-0592.

\end{acknowledgments}


\begin{thebibliography}{99}

\bibitem{Haug}{H. Haug and A.-P. Jauho, {\it Quantum Kinetics in Transport and Optics of 
Semiconductors} (Springer, Berlin, 1996).}

\bibitem{Wiel}{W.G. van der Wiel, S.De Franceschi, J.M. Elzerman, T. Fujisawa, S. Tarucha, L.P. 
Kouwenhoven, Rev. Mod. Phys. {\bf 75}, 1 (2003).}

\bibitem{Holleitner1}{A.W. Holleitner, C.R. Decker, H. Qin, K. Eberl, and R.H. Blick, Phys. Rev. 
Lett. {\bf 87}, 256802 (2001).}

\bibitem{Holleitner2}{A.W. Holleitner, R.H. Blick, A.K. H\"uttel, K. Eberl, and J.P. Kotthaus, 
Science {\bf 297}, 70 (2002).}

\bibitem{ChenJ}{J.C. Chen, A.M. Chang, and M.R. Melloch, Phys. Rev. Lett. {\bf 92}, 176801 
(2004).}

\bibitem{Loss}{D. Loss and E.V. Sukhorukov, Phys. Rev. Lett. {\bf 84}, 1035 (2000).}

\bibitem{Sukhorukov}{E.V. Sukhorukov, G. Burkard, and D. Loss, Phys. Rev. B \textbf{63}, 125315 
(2001).}

\bibitem{Lev}{Lev G. Mourokh, N.J.M. Horing, and A. Yu. Smirnov, Phys. Rev. B {\bf 66}, 85332 
(2002).}

\bibitem{Konig}{J. K\"onig, and Y. Gefen, Phys. Rev. B {\bf 65}, 45316 (2002).}

\bibitem{Kubala}{B. Kubala, and J. K\"onig, Phys. Rev. B {\bf 65}, 245301 (2002).}

\bibitem{Shahbazyan}{T.V. Shahbazyan and M.E. Raikh, Phys. Rev. B {\bf 49}, 17123 (1994).}

\bibitem{Ladron}{M.L. Ladr\'on de Guevara, F. Claro, and P.A. Orellana, Phys. Rev. B {\bf 67} 
195335 (2003).}

\bibitem{Dong1}{Bing Dong, Ivana Djuric, H.L. Cui, and X.L. Lei, J. Phys.: Condens. Matter {\bf 
16}, 4303 (2004).}

\bibitem{Nazarov}{Yu.V. Nazarov, Physica B {\bf 189}, 57 (1993).}

\bibitem{Stoof}{T.H. Stoof and Yu.V. Nazarov, Phys. Rev. B {\bf 53}, 1050 (1996); B.L. Hazelzet, 
M.R. Wegewijs, T.H. Stoof, and Yu.V. Nazarov, Phys. Rev. B {\bf 63}, 165313 (2001).}

\bibitem{Gurvitz}{S.A. Gurvitz and Ya.S. Prager, Phys. Rev. B {\bf 53}, 15932 (1996)}

\bibitem{Gurvitz1} {S.A. Gurvitz, Phys. Rev. B {\bf 57}, 6602 (1998).}

\bibitem{Dong}{Bing Dong, H.L. Cui, and X.L. Lei, Phys. Rev. B {\bf 69}, 35324 (2004).}

\bibitem{Blanter}{Ya.M. Blanter and M. B\"uttiker, Phys. Rep. {\bf 336}, 1 (2000).}

\bibitem{Beenakker}{C. Beenakker and C. Sch\"onenberger, Phys. Today, May 2003, 37 (2003).}

\bibitem{Sun}{H.B. Sun and G.J. Milburn, Phys. Rev. B {\bf 59},
10748 (1999).}

\bibitem{Elattari}{B. Elattari and S.A. Gurvitz, Phys. Lett. A {\bf 292}, 289 (2002).}

\bibitem{Aguado}{R. Aguado and T. Brandes, Phys. Rev. Lett. {\bf 92}, 206601 (2004).}

\bibitem{Djuric}{I. Djuric, Bing Dong, H.L. Cui, {\it IEEE} Transactions on Nanotechnology {\bf 
4}, 71 (2005); Appl. Phys. Lett. {\bf 87}, 032105 (2005).}

\bibitem{Djuric2}{I. Djuric, Bing Dong, H.L. Cui, J. Appl. Phys. {\bf 99}, 63710 (2006).}

\bibitem{Kieblich}{G. Kie\ss lich, P. Samuelsson, A. Wacker, and E. Sch\"oll, Phys. Rev. B, {\bf 
73}, 33312 (2006).}

\bibitem{Kieblich1}{G. Kie\ss lich, A. Wacker, and E. Sch\"oll, Phys. Rev. B, {\bf 68}, 125320 
(2003).}

\bibitem{Aghassi}{J. Aghassi, A. Thielmann, M.H. Hettler, and G. Sch\"on, Phys. Rev. B {\bf 73}, 
195323 (2006).}

\bibitem{Dong4}{Bing Dong and X.L. Lei, cond-mat/0611552 (2006).}

\bibitem{Schwinger}{J. Schwinger, J. Math Phys. {\bf 2}, 417 (1961).}

\bibitem{Ackerhalt}{J.R. Ackerhalt and J.H. Eberly, Phys. Rev. D {\bf 10}, 3350 (1974).}

\bibitem{Cohen}{C. Cohen-Tannoudji, J. Dupont-Roc, and G. Grynberg, {\em Atom-Photon 
Interactions: Basic Processes and Applications}, Wiley, New York, 1992 (Complements Av, pp. 388 
ff and C$_{IV}$, pp. 334 ff).}

\bibitem{Milonni}{P.W. Milonni, {\em The Quantum Vaccum: An Introduction to Quantum 
Electrodynamics}, Academic Press, San Diego, 1994.}

\bibitem{Gardiner}{C.W. Gardiner, P. Zoller, {\em Quantum Noise}, Springer, Berlin, 1999}

\bibitem{Smirnov}{G.F. Efremov and A.Yu. Smirnov, Zh. \'Eksp. Teor. Fiz. {\bf 80}, 1071 (1981) 
[Sov. Phys. JETP {\bf 53}, 547 (1981)].}

\bibitem{Dong2}{Bing Dong, N.J.M. Horing, and H.L. Cui, Phys. Rev. B {\bf 72}, 165326 (2005).}

\bibitem{Korotkov}{A.N. Korotkov, D.V. Averin, and K.K. Likharev, Phys. Rev. B {\bf 49}, 7548 
(1994).}

\bibitem{Mahan}{G.D. Mahan, {\em Many-Particle Physics.} (Third edition, Kluwer Academic/Plenum 
Publisher, New York, 2000).}

\bibitem{Davies}{J.H. Davies, P. Hyldgaard, S. Hershfield, and J.W. Wilkins, Phys. Rev. B {\bf 
46}, 9620 (1992).}

\bibitem{Korotkov1}{A.N. Korotkov, D.V. Averin, K.K. Likharev, and S.A. Vasenko, in {\em 
Single-Electron Tunneling and Mesoscopic Devices}, edited by H. Koch and H. L\"ubbig, Springer 
Series in Electronics and Photonics Vol. 31 (Springer-Verlag, Berlin, 1992), p. 45; A.N. 
Korotkov, Phys. Rev. B {\bf 49}, 10381 (1994).}

\bibitem{Wiseman}{H.M. Wiseman and G.J. Milburn, Phys. Rev. A {\bf 47}, 1652 (1993).}

\bibitem{Datta}{S. Datta, W. Tian, S. Hong, R. Reifenberger, J.I. Henderson, and C.P. Kubiak, 
Phys. Rev. Lett. {\bf 79}, 2530 (1997); V. Mujica, A.E. Roitberg and M.A. Ratner, J. Chem. Phys. 
{\bf 112}, 6834 (2000); S. Pleutin, H. Grabert, G.L. Ingold, and A. Nitzan, J. Chem. Phys. {\bf 
118}, 3756 (2003)}

\bibitem{Pati}{S. Lakshmi and S.K. Pati, Phys. Rev. B {\bf 72}, 193410 (2005); B. Song, D.A. 
Ryndyk, and G. Cuniberti, Phys. Rev. B {\bf 76}, 45408 (2007).}

\bibitem{Pedersen}{J.N. Pedersen, B. Lassen, A. Wacker, and M.H. Hettler, Phys. Rev. B {\bf 75}, 
235314 (2007).}

\bibitem{Tour}{J. Chen, M.A. Reed, A.M. Rawlett, and J.M. Tour, Science {\bf 286}, 1550 (2001).}

\bibitem{Ma}{J. Ma, Bing Dong, and X.L. Lei, Europ. Phys. Jour. B {\bf 36}, 599 (2003).}

\bibitem{MacDonald}{D.K.C. MacDonald, Rep. Prog. Phys. {\bf 12}, 56 (1948).}

\bibitem{Chen}{L.Y. Chen and C.S. Ting, Phys. Rev. B {\bf 46}, 4714 (1992).}

\end{thebibliography}
\end{document}